\documentclass[twocolumn,preprintnumbers,amsmath,amssymb,10pt,a4paper,nofootinbib]{revtex4}
\usepackage{geometry}
\usepackage{hyperref}
\hypersetup{colorlinks=true, citecolor=blue, urlcolor=blue, linkcolor=blue}
\usepackage{mathtools}
\usepackage{graphicx}
\usepackage{dcolumn}
\usepackage{tikz}
\usepackage{bm}
\newcommand{\be}{\begin{equation}}
\newcommand{\ee}{\end{equation}}
\newcommand{\bd}{\begin{displaymath}}
\newcommand{\ed}{\end{displaymath}}
\newcommand{\BE}{\begin{eqnarray}}
\newcommand{\EE}{\end{eqnarray}}

\newcommand{\bx}{\ensuremath{\mathbf{x}}}

\newcommand{\avg}[1]{\left\langle{#1}\right\rangle}

\usepackage{xcolor}

\usepackage{comment}
\begin{document}

\preprint{}
\title{Complex dynamics in the Sherrington--Kirkpatrick game}

\author{Desmond Chan}
\affiliation{Department of Informatics, King's College, London WC2B 4BG, UK}
\email{desmond.chan@kcl.ac.uk}
 
\author{Tobias Galla}
 \affiliation{Instituto de F\'isica Interdisciplinar y Sistemas Complejos,IFISC, Edifici Instituts Universitaris de Recerca,
Campus Universitat de les Illes Balears
E-07122 Palma, Mallorca, Spain}
\email{tobias.galla@ifisc.uib-csic.es}

\date{\today}

\begin{abstract}
We study the outcome of adaptive learning of a large number of players engaging in sets of two-strategy two-player games. We are interested in typical games, and generate the payoff matrices at random at the beginning. The payoff matrices then remain fixed during the learning process. This provides a game theoretic foundation for the Sherrington-Kirkpatrick (SK) game, recently introduced by Garnier-Brun, Benzaquen and Bouchaud. The original model by these authors is a special case, with no bias towards any strategy. We here determine stability of learning for SK games with general random bias, and find that the nature of the stable state is affected by random fields. We also introduce a grand-canonical version of the SK game, in which players can choose to abstain. We determine the stability of learning for this game. Our analysis confirms that complex situations involving many players are frequently unlearnable, even if each player only chooses between two different actions. The rate with which players lose memory of past payoffs and the competitiveness of the game emerge as key parameters determining whether learning converges to a unique fixed point, whether there are many fixed points, or if the dynamics remains persistently volatile. 
\end{abstract}
\maketitle

\section{Introduction}
While there is no unique definition of a complex system, complexity is generally understood to involve the emergence of unexpected and perhaps unpredictable behaviour from the interaction of relatively simple agents. What these agents represent depends on the context. They could be members of a population in a model of the spread of an epidemic, genes and proteins in developmental biology, or traders in a financial market. Many of these systems involve competition for limited resources, strategic decision making and adaptation. Since its inception in the 1940s \cite{vonNeumann1944}, game theory has been an important ingredient of modelling these processes. There has been significant interest in the physics community in evolutionary games in particular, and in the emergence of cooperation and the general phenomenology of different games, see for example \cite{hauert2005game, traulsen2009stochastic}.

The bulk of existing work in game theory, both in physics and in other disciplines, focuses on simple games. By this we mean games with a small number of players who each only have a small number of `actions' (pure strategies) at their disposal. Classical examples are the prisoners' dilemma, or cyclic games such as rock-paper-scissors. In these games only two players interact, and they choose between two actions in the prisoner's dilemma, or three actions in the rock-paper-scissors game.

Much of the modelling work on strategic decision making in economics is based on the assumption of perfectly rational players who have full information about the game, and who assume that all other players are also fully rational and fully informed. Under these assumptions, the natural outcomes of a game are the so-called Nash equilibria \cite{Nash}. These are points in strategy space such that no player can improve their payoff by {\em unilaterally} changing their strategy. Nash proved that such points always exist in sufficiently simple games -- broadly when the number of players is finite and when each player only has a finite number of actions to choose from. Importantly, Nash equilibria can be probabilistic combinations of pure strategies. For example the Nash point for the rock-paper-scissors game is given by the fully mixed strategy, in which each player chooses each of the three actions randomly with probability $1/3$ each.  

For simple games such as rock-paper-scissors, the concept of a Nash equilibrium is perfectly sensible, as the equilibrium points are often unique and they are easy to find. In more complex situations it is more questionable if players would be able to find Nash equilibria. In some complicated games there can be many Nash points \cite{berg1,berg2,mclennan}, and algorithms to find these equilibria typically require a computing time that grows exponentially with the size of the game \cite{daskalakis}.  

With this in mind it is not surprising that Nash equilibria are not the natural outcome in models of human players learning to play a game. Sato and co-workers for example have shown that learning can lead to chaotic dynamics even in low-dimensional games \cite{sato1,sato2,sato3}. This casts doubt on the usefulness of the idea of Nash equilibria in more complex games.

Complexity in normal-form games can arise in (at least) two distinct ways: (i) by increasing the number of actions available to each player while keeping the number of players small, or (ii) by increasing the number of players while keeping the action space small.

The first route was taken for example in \cite{galla2013complex, sanders2018prevalence, pangallo2019best}. Building on earlier work on high-dimensional replicator models \cite{opper1992phase}, Galla and Farmer \cite{galla2013complex} used simulations and analytical methods from the theory of disordered systems to demonstrate different dynamical outcomes for experience-weighted attraction learning in complicated two-player games. They studied the learning dynamics of two players who engage in repeated game with many actions and where the payoff matrix is drawn at random at the beginning. One main result is that the learning process remains chaotic and unpredictable for a large range of games. This was later extended to multi-player games \cite{sanders2018prevalence}, where chaos is even more prevalent. Complexity in these studies resulted mainly from a large strategy space.

To investigate  scenario (ii) -- situations with finite action spaces, but a larger number of players -- Garnier-Brun, Benzaquen, and Bouchaud \cite{garnier2024unlearnable} introduced the `Sherrington--Kirkpatrick game' (SK game). In this game, a large number of players repeatedly interact, each choosing between two possible actions. Similar to \cite{galla2013complex}, the interaction between agents is governed by random interaction coefficients, drawn at the beginning and then fixed as the learning process unfolds. As shown in \cite{garnier2024unlearnable} learning in this game shows rich collective dynamics, and the players are generally unable to learn the best-possible point in strategy space. This leads to an outcome based on the idea of `satisficing' -- players must make do with sub-optimal rewards but which are nonetheless satisfactory and sufficing. 

The term `Sherrington--Kirkpatrick game' highlights similarities with the celebrated Sherrington--Kirkpatrick model of spin glass physics \cite{sherrington1975solvable}. In the SK game players choose between two actions (akin to spin up or down in the SK spin glass model), and as in the SK spin glass model there are quenched random interaction coefficients. In this context, we also note work by Menezes and Sherrington on Minority Games with SK interactions \cite{menezes2013minority}, where there are also many players, each with only two actions to choose from.

The choice to study ensembles of random payoff matrices  both in \cite{galla2013complex, sanders2018prevalence, pangallo2019best} and in \cite{garnier2024unlearnable} follows the ideas first proposed for example by May \cite{may1972will} to characterise large complex systems. Such systems will generally have many interaction coefficients (typically if there are $N$ degrees of freedom, there would be order $N\times N$ coefficients). It would be impossible to study the outcome in detail as a function of so many input parameters. Additionally, this would not provide much useful insight either. Instead May therefore introduced a statistical view, and determined outcome as a function of key summary statistics of the interaction coefficients, such as for example their variance.  In the context of the learning of complex games, this means to look for key parameters of the ensemble of random games and the learning algorithm that determine if and when learning converges, or if remains volatile. Depending on how competitive the game is and on how quickly players adapt, the authors of \cite{galla2013complex} and \cite{sanders2018prevalence} found regimes for two-player games with many actions in which learning converges to a unique stable fixed point (albeit not a Nash equilibrium), phases in which there are many stable fixed points, and a regime in which learning remains chaotic and unpredictable. Garnier-Brun et al \cite{garnier2024unlearnable} find broadly similar phases, additionally they also investigated cyclic or quasi-cyclic behaviour in the SK game.

In this paper, we extend the work of \cite{garnier2024unlearnable} in several different ways: We provide a bottom-up construction of the SK game dynamics starting from concrete payoff matrices. The game studied in \cite{garnier2024unlearnable} is in fact a special case of a more general setup. More specifically, Garnier-Brun et al focus on a model without external field. This describes a setup in which no player has any intrinsic bias to any of their two strategies. As we will describe, some of the phenomena reported in \cite{garnier2024unlearnable} are restricted to that special case, and allowing for non-zero random fields changes the behaviour and phase diagram. We also introduce a version of the SK game in which some players can choose whether or not to participate in the game at any one time step, but where other players must participate at each step. Thus, the number of active players varies in time. We call this the `grand-canonical SK game', in analogy with the grand-canonical ensemble in statistical physics and with the so-called `grand-canonical minority game' \cite{challet2006minority}. 

The outline of the paper is as follows: In Sec.~\ref{sec:def} we describe how the SK game can be constructed from a concrete payoff structure, and how this leads to setups with and without random fields. Section \ref{sec:gf} contains a formal analysis, using dynamic mean field theory -- a well-established technique in disordered systems. In Sec.~\ref{sec:stab} we discuss and compare the predictions from this theory with numerical simulations. In particular we establish the different types of behaviour in the games without and with field, respectively. Section \ref{sec:gcsk} contains the definition and analysis of the grand-canonical SK model. We highlight how this model differs from the SK game in which all players must participate at each step. We present our conclusions in Sec.~\ref{sec:concl}.

\section{SK Model formulation}\label{sec:def}
\subsection{Mixed strategies and payoff structure}
We consider a setup with $N$ players. We label the players by letters $i$ or $j$ in $\{1,\dots,N\}$. In the SK game, each player has two pure strategies  (`actions') to choose from, we label these actions $\pm 1$. The game proceeds in discrete rounds. We write $s_i(t)=1$ if player $i$ takes action $1$ at time $t$, and $s_i(t)=-1$ if they play action $-1$. The variable $x_i(t)$ denotes the mixed strategy of player $i$ at time $t$. More precisely $x_i(t)$ is the probability for player $i$ to play action $+1$ at that time. They play action $-1$ with probability $1-x_i(t)$. We write $\bx=(x_1,\dots,x_N)$.
 
The interaction between players occurs through two-player games. These games are `simple' $2\times 2$ games, in the sense that only two players participate in each game, and they each only have two actions to choose from. The complexity in the model arises from the assumption that each player engages in a large number of two-player games with different opponents.

More specifically, in each round each player plays one two-player game with each of the other players. Each player decides what action to take at any given time step, and that same action $s_i(t)=\pm 1$ is then applied to all two-player games with the other players. In the next step the player can take a different action. Each player derives a payoff from these interactions, via a set of payoff matrices, which we will define below. This setup is combined with a learning dynamics that the players use to update their mixed strategies. This will also be described in more detail below.

We write $a^{++}_{ij}$ for the payoff player $i$ receives from the interaction with player $j$ if $i$ plays action $+1$, and $j$ plays $+1$. Similarly, $a^{+-}_{ij}$ is the payoff to $i$ if $i$ plays $+1$ and $j$ plays $-1$. The quantities $a^{-+}_{ij}$ and $a_{ij}^{--}$ are defined analogously.

Throughout the paper, we assume that the $a_{ij}^{ss'}$ ($s,s'\in\{\pm 1\}$) are drawn randomly at the beginning, and then remain fixed as the dynamics unfold. Thus, the payoff matrices are the quenched disorder of the problem. We will define the precise ensemble from which these payoff matrix elements are drawn further below.
 
 \subsection{Sato--Crutchfield learning}
 In the model each player makes the decision what action to play based on a pair of propensities. These describe a tendency for the player to play either action. We label these $q_i^\pm(t)$, such that $q_i^+(t)$ is the propensity for player $i$ to play action $+1$ at time $t$, and similarly for $q_i^-(t)$. As described below, these propensities capture the outcome of the player's learning process. The propensities determine the player's mixed strategy via a softmax (or logit) rule
 \be\label{eq:xpm}
 x_i^\pm(t)=\frac{e^{\beta q_i^\pm(t)}}{e^{\beta q_i^+(t)}+e^{\beta q_i^-(t)}}.
 \ee
  By construction $x_i^+(t)+x_i^-(t)=1$ for all $i$ and $t$. 
 
 The coefficient $\beta\geq 0$ is a parameter of the learning and decision model, it describes an `intensity of choice'. We assume that all players use the same intensity of choice. For very large $\beta$, each player is very likely to play the action with the higher propensity. For $\beta=0$ instead, the player plays each of the two actions with probability $1/2$, irrespective of the propensities.

 To simplify the notation, we introduce the following quantities
 \BE\label{eq:Pi_pm}
 \Pi_i^+[\bx]&=&\sum_j a_{ij}^{++}x_j^++\sum_j a_{ij}^{+-}x_j^-, \nonumber \\
  \Pi_i^-[\bx]&=&\sum_j a_{ij}^{-+}x_j^++\sum_j a_{ij}^{--}x_j^-.
\EE
We have here written $x_j^+=x_j$ for the probability of player $j$ to use action $+1$, and $x_j^-=1-x_j$ for the player to play action $-1$. The quantity $\Pi_i^s[\bx]$ is the (expected) payoff to player $i$ if $i$ plays action $s\in\{\pm1\}$, and given the mixed strategies $\{x_j^\pm\}$ of the remaining players. We use the notation $\bx=(x_1^+,\dots, x_N^+)$. We will occasionally omit the argument $\bx(t)$ of the payoffs, to keep the notation compact.

 To update the propensities the players use Sato--Crutchfield learning \cite{sato1, sato2, sato3}, a variant of Experienced-Weighted Attraction learning (EWA)  \cite{camerer1999experience}. More specifically, we use a batch-learning model \cite{sato3, galla2013complex}. This assumes that players take an average over a large (formally infinite) batch of successive interactions with the other players between updates of their propensities. The updates are carried out assuming that no player changes their mixed strategy within a batch of interactions. The update rules then are
    \BE\label{eq:batch_2}
 q_i^+(t+1)&=&(1-\alpha)q_i^+(t)+ \Pi_i^+[\bx(t)], \nonumber \\
 q_i^-(t+1)&=&(1-\alpha)q_i^-(t)+ \Pi_i^-[\bx(t)].
 \EE 
Thus, the update consists of two components. The first is the addition of $\Pi_i^\pm[\bx(t)]$ to the respective propensities $q_i^\pm$. In simple terms, agent $i$ adds to their propensity for action $+1$ the average payoff they would have received in a large batch of games, if they (player $i$) had always played action $+1$, and given the other players' current mixed strategies. Similarly, $\Pi_i^-[\bx(t)]$ is added to $q_i^-$.

The second component of the update is memory loss, characterised by the parameter $\alpha\in[0,1]$. For $\alpha=0$, the update simply accumulates payoffs. The propensities $q_i^\pm(t)$ are then the total (expected) payoffs player $i$ would have received had they always played action $\pm 1$, respectively, up to time $t$. For $\alpha=1$ on the other hand, past rounds are immediately forgotten, and only the most recent batch is taken into account to compute propensities for the next time step. Intermediate values of $\alpha$ describe exponential discounting of past payoffs, with a time scale $1/\alpha$. Different choices for the learning dynamics are possible, for example the authors of \cite{garnier2024unlearnable} place a factor of $\alpha$ in front of the term $\Pi_i^\pm$ in Eq.~(\ref{eq:batch_2}). Except for special cases, this can be absorbed into a re-definition of the intensity of choice $\beta$, and makes no major difference. We follow the conventions of \cite{galla2013complex}.

Combining Eqs.~(\ref{eq:batch_2}) and (\ref{eq:xpm}) we have
 \be\label{eq:x_update}
 x_i^\pm(t+1)=\frac{x^\pm_i(t)^{1-\alpha}e^{\beta \Pi_i^\pm[\bx(t)]}}{x^+_i(t)^{1-\alpha}e^{\beta \Pi_i^+[\bx(t)]} + x_i^-(t)^{1-\alpha}e^{\beta \Pi_i^-[\bx(t)]}}.
  \ee

Given $x_i^+(t)+x_i^-(t)=1$ for all $i,t$, we can reduce the dynamical system to one variable per player. We choose these variables as
 \be
 m_i(t)=x_i^+(t)-x_i^-(t),
 \ee
taking values in the interval $-1\leq m_i\leq 1$. If $m_i(t)=-1$ then player $i$ plays action $-1$ with probability one at time $t$. If $m_i(t)=1$, they play action $+1$. For $m_i(t)=0$ the player plays each action with probability $1/2$. 

Following \cite{galla2013complex} we take the continuous-time limit $\alpha,\beta\to 0$ at fixed ratio $\lambda=\alpha/\beta$. As detailed in Appendix~\ref{sec:eq_for_m} we then find

 \BE\label{eq:m_learning}
 \frac{d}{dt} m_i = \frac{1}{2}(1-m_i^2)\bigg [\Pi_i^+- \Pi_i^- - \lambda \ln\frac{1+m_i}{1-m_i}\bigg].
 \EE
 The right-hand side is proportional to $1-m_i^2$, this makes sure $m_i$ never leaves the interval $[-1,1]$. The first term in the square bracket is akin to what one would see in a replicator equation. If $\Pi_i^+>\Pi_i^-$ then this term promotes an increase of $m_i$, and thus player $i$ becomes more likely to play action $+1$. For $\Pi_i^+<\Pi_i^-$, this term leads to a reduction of $m_i$, making action $-1$ more likely. 
 
 The term proportional to $\lambda$ represents memory loss. Recalling that $\lambda\geq 0$, it acts as a force towards a random strategy $m_i=0$ where the player plays each of the two actions with probability $1/2$. For example, when $m_i$ approaches one from below, the expression $- \lambda \ln\frac{1+m_i}{1-m_i}$ will become large and negative, and dominate over the payoff term $\Pi_i^+-\Pi_i^-$. This will lead to a reduction of $m_i$.
 \subsection{The Sherrington--Kirkpatrick game}
We now investigate the payoff structure in more detail and show how exactly the dynamics in \cite{garnier2024unlearnable} can be constructed.

We introduce 
 \be
 A_{ij}^+ = a_{ij}^{++}-a_{ij}^{-+}, \qquad A_{ij}^-=a_{ij}^{+-}-a_{ij}^{--}.
 \ee
The quantity $A_{ij}^+$ is the improvement in payoff player $i$ can achieve in the interaction with $j$ by playing strategy $+1$ as compared $i$ to playing $-1$, and assuming that $j$ plays action $+1$. The coefficient $A_{ij}^-$ is a similar payoff difference, but now assuming $j$ plays $-1$. 
 
 The time evolution of the $m_i$ in Eq.~(\ref{eq:m_learning}) only depends on payoff differences, which can be written as follows, 
  \BE\label{eq:pi_diff}
  \Pi_i^+-\Pi_i^- &=& \frac{1}{2}\sum_j (A_{ij}^+ - A_{ij}^-)m_j+\frac{1}{2}\sum_j (A_{ij}^++A_{ij}^-).\nonumber \\
 \EE
The term containing $A_{ij}^++A_{ij}^-$ in this equation acts as a random field. A non-zero value of this quantity indicates that one of player $i$'s actions does intrinsically better than the other action when playing against $j$, and when the other player $j$ acts randomly ($x_j^\pm = 1/2$, i.e. $m_j=0$).  We now distinguish between two different scenarios, leading to dynamics without and with such random fields, respectively. In the following, when we speak of `a game' we always mean the collection of coefficients $A_{ij}^+$ and $A_{ij}^-$ for all pairs $i,j$.

 \subsubsection{Single-matrix game, no external field}
We first focus on games where there is no background advantage to any action. This setup that leads to the dynamics as described in \cite{garnier2024unlearnable}. 

The background bias vanishes when $A_{ij}^+=-A_{ij}^-$ for all $i,j$, and we will assume this is the case throughout this subsection. We simply write $A_{ij}$ for this quantity. We then have the following picture: If player player $j$ plays action $+1$ then player $i$ scores $A_{ij}$ more points by playing $+1$ as compared to player $i$ using action $-1$. If player $j$ plays $-1$, then player $i$ scores $A_{ij}$ points more by playing $-1$ as compared to playing $+1$. A similar interpretation holds for $A_{ji}$. 

This can be illustrated as follows (we draw attention to a similar table in the Appendix of \cite{garnier2024unlearnable})
\be\label{eq:diff_table1}
\begin{tabular}{c|cc}
~ &~~ \mbox{$j$ plays +1}~~ & ~~\mbox{$j$ plays -1}  \\
\hline
\mbox{$i$ plays $+1$} & $A_{ij}/2$ & $-A_{ij}/2$\\
\mbox{$i$ plays $-1$} & $-A_{ij}/2$ &$ A_{ij}/2$,
\end{tabular}
\ee 
where the entries illustrate the payoff to player $i$. We stress that this is not an actual payoff matrix, but is only accurate as far as payoff {\em differences} go (payoff to player $i$ when $i$ plays action $+1$ compared to $i$ playing action $-1$). Player $j$ faces a similar table
\be
\begin{tabular}{c|cc}
~ & \mbox{$i$ plays +}~~ & \mbox{$i$ plays -}  \\
\hline
\mbox{$j$ plays $+$} & $A_{ji}/2$ & $-A_{ji}/2$\\
\mbox{$j$ plays $-$} & $-A_{ji}/2$ &$ A_{ji}/2$.
\end{tabular}
\ee 
The game is now entirely defined by the $N\times N$ matrix $A_{ij}$. It is important to remember that unlike in \cite{galla2013complex} $i$ and $j$ label players, not actions.

 The model we end up with has payoff difference 
$\Pi_i^+-\Pi_i^- = \sum_j A_{ij} m_j$, see Eq.~(\ref{eq:pi_diff}),
  and hence a dynamics given by (after re-scaling time to absorb a factor of $1/2$),
 \be\label{eq:dyn_m}
 \frac{d}{dt} m_i = (1-m_i^2) \left(\sum_j A_{ij} m_j - \lambda \ln\frac{1+m_i}{1-m_i}\right).
\ee
This is the continuous-time analog of Eq. (7) in \cite{garnier2024unlearnable}, where we recall again the different conventions regarding the memory-loss parameter. A brief further discussion can also be found in Appendix~\ref{app:compare}.

We next discuss correlations between $A_{ij}$ and $A_{ji}$. Imagine a situation where $A_{ij}=A_{ji}$ for a particular pair of players. If $A_{ij}=A_{ji}>0$, and setting the other plays aside, the best outcome for the pair $i,j$ occurs if they both play $+1$, or both $-1$. If $A_{ij}=A_{ji}<0$, then the best outcome for both players is to take opposite actions (one player plays $+1$, the other $-1$). So as far as the interaction between $i$ and $j$ is concerned, there is no competition when $A_{ij}=A_{ji}$. If one of the mutually favourable states has been reached neither player can unilaterally increase their payoff from the interaction with this particular other player.

Assume now instead a situation in which $A_{ij}=-A_{ji}$, and $A_{ij}>0$. Player $i$ will then always want to do what player $j$ does. Player $j$ on the other hand will want to do the opposite of what $i$ does. The roles of $i$ and $j$ are reversed if $A_{ij}=-A_{ji}<0$. This is akin to the matching pennies game (see e.g., \cite{Traulsen2004PRL}). Whatever combination of actions is played by $i$ and $j$ ($++, +-, -+$, or $--$) there is always one player who can do better by unilaterally changing their action. Also, the two payoffs (that to $i$ and that to $j$) always sum to a constant (keeping in mind that the matrices above are not actually payoff matrices). Thus, up to a constant background payoff, this is a zero sum game.

\medskip

As motivated in the introduction, our work does not focus on particular instances of the payoff matrices. Instead, we focus on the limit of a large number of players (formally, $N\to \infty$), and our goal is to characterise the outcome of learning in terms of the statistics of the payoff matrix entries. We therefore assume that the $A_{ij}$ are drawn from a joint distribution, and treat the first and second moments as control parameters.  Maximum-entropy arguments \cite{galla2013complex} then lead to a Gaussian distribution. In order not to introduce any bias, we assume that the mean of each $A_{ij}$ vanishes. Considering the limiting cases $A_{ij}=A_{ji}$ and $A_{ij}=-A_{ji}$ discussed above, we allow for correlations between $A_{ij}$ and $A_{ji}$. We then arrive at a Gaussian random matrix ensemble with
\be
\overline{A_{ij}^2}=\overline{A_{ji}^2}=\frac{\sigma^2}{N-1},\qquad \overline{A_{ij}A_{ji}}=\frac{\Gamma \sigma^2}{N-1}
\ee
 for all pairs $i<j$. The overbar denotes averages over the random matrix ensemble. The variances of the matrix elements scales as $(N-1)^{-1}$, this guarantees a well defined analytical limit for $N\to\infty$, and allows us to carry out a statistical physics analysis similar to \cite{galla2013complex, garnier2024unlearnable}. We choose a factor of $N-1$ here because, players are assumed not to play against themselves. In the limit $N\gg 1$ it makes no difference whether the variance scales as $(N-1)^{-1}$ or as $N^{-1}$, as is more common in the spin glass literature. 

 We note that the parameter $\sigma$ sets the scale of the payoffs, and given that payoffs are multiplied by the strength of selection [Eq.~(\ref{eq:x_update})], only the product of $\beta$ and $\sigma$ is relevant. Equivalently, any changes of $\sigma$ can be absorbed into changes of the parameter $\lambda=\alpha/\beta$. We use $\sigma=1$ throughout in all figures. 
 
 The parameter $\Gamma$ takes values in the interval from $-1$ to $1$. When $\Gamma=-1$, we have $A_{ij}=-A_{ji}$ with probability one. This is akin to a zero-sum game. For $\Gamma=1$ on the other hand, we have $A_{ij}=A_{ji}$. This resembles a potential-game, where the incentives of the players are aligned with a common underlying potential. 
 Values $-1<\Gamma<1$ interpolate between these two extremes.
\subsubsection{Bi-matrix version, random fields}
We can also consider scenarios in which we do not impose $A_{ij}^+ = - A_{ij}^-$ in Eq.~(\ref{eq:pi_diff}). In this situation there can be an intrinsic bias to one of the two actions of player $i$ in the interaction with $j$. The quantity $B_{ij}= (A_{ij}^+ + A_{ij}^-)/2$ quantifies this bias (with positive values representing a bias towards action $+1$). Writing $A_{ij}=(A_{ij}^+-A_{ij}^-)/2$, this can be illustrated in the payoff matrix for player $i$
\be\label{eq:diff_table2}
\begin{tabular}{c|cc}
~ &~~ \mbox{$j$ plays +1}~~ & ~~\mbox{$j$ plays -1}  \\
\hline
\mbox{$i$ plays $+1$} & $(A_{ij} +B_{ij} )/2$ & $ (-A_{ij} + B_{ij})/2$\\
\mbox{$i$ plays $-1$} & $(-A_{ij}-B_{ij} )/2$ &$ (A_{ij} -B_{ij} ) /2$.
\end{tabular}
\ee 
As before, the matrix in \eqref{eq:diff_table2}, like the one in
\eqref{eq:diff_table1}, should be understood as encoding only the payoff
differences for player $i$ that arise from changing their own action.
More precisely, suppose first that player $j$ plays $+1$. Then the payoff
advantage to player $i$ from playing $+1$ rather than $-1$ is given by:
$A_{ij}+B_{ij}
    = A_{ij}^{+}.$
Similarly, if player $j$ plays $-1$, the payoff advantage to player $i$ from
playing $+1$ rather than $-1$ is given by:
$-A_{ij}+B_{ij}
    = A_{ij}^{-}.$

The learning dynamics in Eq.~(\ref{eq:dyn_m}) can now be written as:
   \BE \label{eq:modified_dyn_m}
 \frac{d}{dt} m_i &=& (1-m_i^2) \bigg\{ - \lambda  \ln\frac{1+m_i}{1-m_i}+\sum_j A_{ij} m_j + h_i \bigg\}, \nonumber \\
 \EE
 where $h_i = \sum_j B_{ij}$.
 
Similar to $A_{ij}$, we take the $B_{ij}$ to be quenched Gaussian random numbers with zero mean, and with 
\begin{align}
   \overline{(B_{ij})^2 }  =  \frac{\Delta^2}{N-1},  
\end{align}
where $\Delta$ controls the variances of $B_{ij}$. For $\Delta=0$ we return to Eq.~(\ref{eq:dyn_m}). In this limit, there is no intrinsic bias for any player to choose any given action. A positive value of $\Delta$ means each player has a random intrinsic bias towards a given action. As $\Delta^2$ becomes large (compared to $\sigma^2$) players become increasingly biased towards their favourite actions, and in the limit $\Delta^2\to\infty$,  they always play this action. 

We could also allow the quantities $B_{ij}$ and $B_{ji}$ to be correlated, with
$
\overline{B_{ij}B_{ji}}=\frac{\gamma \Delta^2}{N-1}.
$
However, the dynamic mean field calculation shows that correlations between $B_{ij}$ do not appear in the effective single-player description. As a result, it does not affect the behaviour of the system when the number of players becomes very large, $N\to\infty$.
We also note that $A_{ij}$ and $B_{ij}$ are uncorrelated. This is not an extra assumption, but follows directly from how these two quantities are defined.

\section{Generating functional analysis}\label{sec:gf}
\subsection{Effective single-player dynamics}
A dynamic mean field theory for the SK games with and without field can be derived following standard generating-functional methods \cite{de_dominicis}, see also \cite{coolen2005mathematical, galla2024generating}. Alternatively, the cavity method can be used (see e.g. \cite{bunin2017, roy2019numerical}). We do not include the full derivation in the main paper, but a summary can be found in Appendix \ref{app:gfsk}. We find the following process for an effective representative player is given by:
\BE\label{eq:eff_proc}
\dot m &=& -\lambda (1-m^2) \ln \frac{1+m}{1-m}\nonumber \\
&&+(1-m^2)\left[\Gamma\sigma^2 \int dt'\, G(t,t') m(t') +\eta(t)\right]. \nonumber \\
\EE
In this expression, the variable $\eta(t)$ is Gaussian coloured noise, and the integral term represents retarded self-interaction. These contributions arise from the average over the disorder. We have the following self-consistency relations
\BE\label{eq:sc_field}
C(t,t')&\equiv& \avg{\eta(t)\eta(t')} = \sigma^2\avg{m(t) m(t')} +\Delta^2, \nonumber \\
G(t,t')&=& \frac{\delta}{\delta \eta(t')} \avg{m(t)}.
\EE
The angle brackets $\avg{\cdots}$ represent an average over realisations of this effective process. Eqs. (\ref{eq:eff_proc}) and (\ref{eq:sc_field}) define a self-consistent problem for the dynamic order parameters $C(t,t')$ and $G(t,t')$. The statistics of the effective process (\ref{eq:eff_proc}) in trajectory space reproduces the statistics of the player-specific paths $m_i(t)$ in the original game learning dynamics.

\subsection{Fixed-point analysis}\label{sec:fp}
We now broadly follow the ideas of \cite{opper1992phase}. Assuming the dynamics converges to a stable fixed point (independent of initial conditions), we have a well-defined solution for the macroscopic order parameters of the form $C(t,t') = q$ and $G(t,t') = G(t-t')$, with causality implying $G(t,t') = 0$ for $t' > t$. We define $\chi = \int_0^\infty d\tau\, G(\tau)$. At the fixed point, $m$ and $\eta$ are static random variables with $\langle m^2 \rangle = q$ and $\langle \eta^2 \rangle = \sigma^2 q + \Delta^2$.

Writing $\eta = \sqrt{q\sigma^2 + \Delta^2}\, z$, where $z \sim \mathcal{N}(0,1)$, the fixed-point equation becomes
\begin{equation}\label{eq:fp}
(1 - m^2)\left[-\lambda \ln \frac{1+m}{1-m} + \Gamma \sigma^2 \chi m + \sqrt{q\sigma^2 + \Delta^2}\, z \right] = 0.
\end{equation}
Besides the boundary solutions $m = \pm 1$, a nontrivial solution satisfies
\begin{equation}\label{eq:m_z}
-\lambda \ln \frac{1+m}{1-m} + \Gamma \sigma^2 \chi m + \sqrt{q\sigma^2 + \Delta^2}\, z = 0.
\end{equation}
Given that $\lambda > 0$, the term $ \lambda \ln \frac{1+m}{1-m} - \Gamma \sigma^2 \chi m$ diverges to $-\infty$ as $m \downarrow -1$ and to $+\infty$ as $m\uparrow 1$, ensuring at least one solution $m \in (-1,1)$ for any given $z$. There are also solutions $m=\pm1$. These correspond to diverging propensities ($m\to \pm \infty$), which is unphysical for any non-zero memory loss parameter ($\lambda>0$). The physically meaningful solution $m(z)$ of Eq.~(\ref{eq:m_z}) is therefore strictly in the interval from $-1$ to $1$, and defines a one-to-one relation between $z$ and $m$.
\footnote{There can be multiple solutions of Eq.~(\ref{eq:m_z}) in the interval $(-1,1)$ in certain circumstances. This arises when $\Gamma > 0$ and $\lambda < 2 \Gamma \sigma^2 \chi$. Although we cannot formally demonstrate that this never occurs for the parameter ranges considered in this work, we have not encountered any multi-valuedness. We therefore hypothesise that for any combination of parameters in the stable fixed-point regime there is a unique solution, $m(z)$, of Eq.~(\ref{eq:m_z}). }

The corresponding self-consistency equations are given by:
\BE\label{eq:self_consistency}
q &=& \int Dz\, m(z)^2, \nonumber \\
\chi &=& \frac{1}{\sqrt{q\sigma^2 + \Delta^2}} \int Dz\, \frac{\partial m(z)}{\partial z},
\EE
where $Dz = \frac{dz}{\sqrt{2\pi}} e^{-z^2/2}$. \\

We now briefly comment on the model without external fields. Our model then comes closest to the work in  \cite{garnier2024unlearnable} where a fixed-point regime with all $m_i=0$ was reported (a `paramagnetic state'). We also find such a state in our model when there are no external fields. Specifically, the fixed point solution to Eq.~(\ref{eq:fp}) is then given by $m(z) = 0 $ for all $z$, and we have $q=0$. All players play fully mixed strategies, i.e., they each play actions $\pm1$ randomly with probability $1/2$ each. The quantity $\chi$ as given in Eq.~(\ref{eq:self_consistency}) is then to be understood as follows. We write $m(\eta) $ for the solution of 
\be\label{eq:m_eta}
-\lambda \ln \frac{1+m}{1-m} + \Gamma \sigma^2 \chi m + \eta = 0,
\ee
and, noting that all probability for $\eta$ concentrates on $\eta=0$ at the paramagnetic solution, we then have $\chi=\left. \frac{\partial m(\eta)}{\partial\eta}\right|_{\eta=0}$. 

As we will demonstrate below, this paramagnetic solution specific to the SK game with no intrinsic bias (external field). The fixed-point phase is more involved when $\Delta^2\neq 0$.  
\subsection{Linear-stability Analysis}\label{sec:lin_stab_main}
Following the steps described in \cite{opper1992phase, roy2019numerical, galla2024generating} we can identify the onset of a linear instability from the mean-field process. Details are given in the appendix. We find that the stability condition can be written as
\begin{align}\label{eq: stab_relation_main}
      \sigma^2 \left\langle \left( \frac{\partial m } { \partial \eta} \right) ^2 \right \rangle < 1.
\end{align}
We comment briefly on the case $\Delta=0$. The self-consistent solution is then $q=0$, and $m\equiv 0, \eta\equiv 0$. Nonetheless, the object $\partial m/\partial \eta$ [with $m(\eta)$ the solution of Eq.~(\ref{eq:m_eta})] is well defined, and will generally be non-zero.

If there is no external field ($\Delta=0$), the stability condition Eq.~(\ref{eq: stab_relation_main}) simplifies to $\sigma < 2\lambda -  \Gamma\sigma^2 \chi$. One can then show (see again the appendix) that the onset of instability occurs when $\lambda$ takes the value
\be
\lambda_\text{crit} = (1+ \Gamma) \sigma/2,
\ee
with instability for $\lambda < \lambda_\text{crit}$. 

\section{Stability diagram and dynamical behaviour}\label{sec:stab}
\subsection{Different dynamical regimes}
\begin{figure}[h]
\centering
\includegraphics[width=0.85\linewidth]{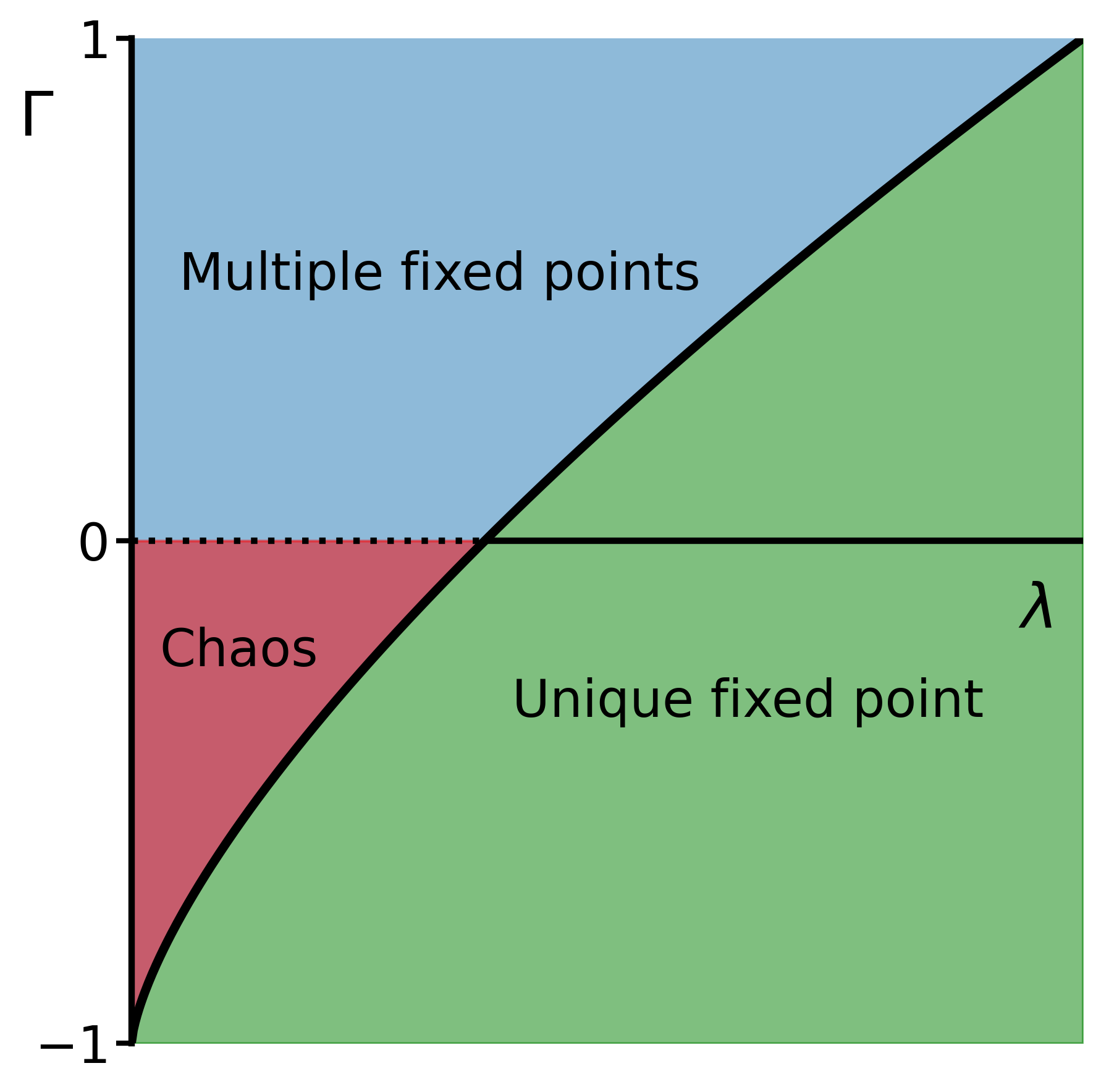}
\caption{Illustration of the general structure of the stability diagram of the SK game as a function of the competition parameter $\Gamma$ and the memory-loss parameter $\lambda$. The phase diagram for the grand-canonical SK game (to be defined below), also has this general structure.}
\label{fig:phase_diag}
\end{figure}

Similar to the complex two-player games with many actions per player in \cite{galla2013complex}, the SK game has three dynamical regimes: (i) When $\lambda$ is large, one finds a phase with a unique stable fixed point for fixed realisations of the payoff matrices; (ii) We find a volatile regime for small $\lambda$ and when $\Gamma < 0$; and finally (iii) a multiple-fixed point regime for small $\lambda$ when $\Gamma > 0$. Figure \ref{fig:phase_diag} shows an illustration of this general structure. We stress that this indicates the dominant behaviour that we expect for very large (infinite) systems in each of the phases. In simulations (necessarily conducted at finite $N$), it is possible that some samples in the formally stable phase remain volatile, or that convergence is found for some runs in the `chaotic' phase. However, these effects generally reduce with an increased number of players. 

\begin{figure}[t]
\centering
     \includegraphics[width=\linewidth]{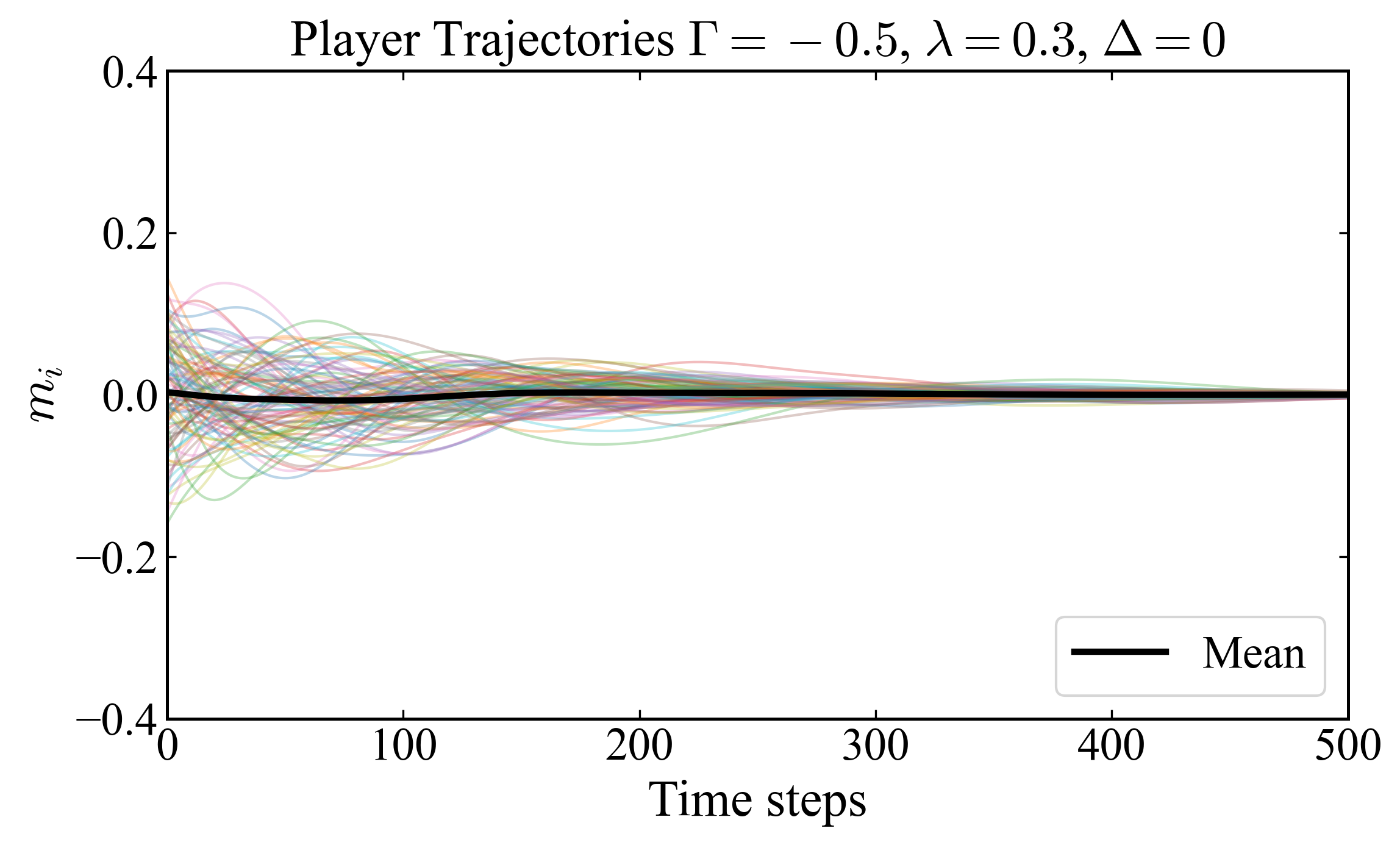}
     
\vfill
  \includegraphics[width=\linewidth]{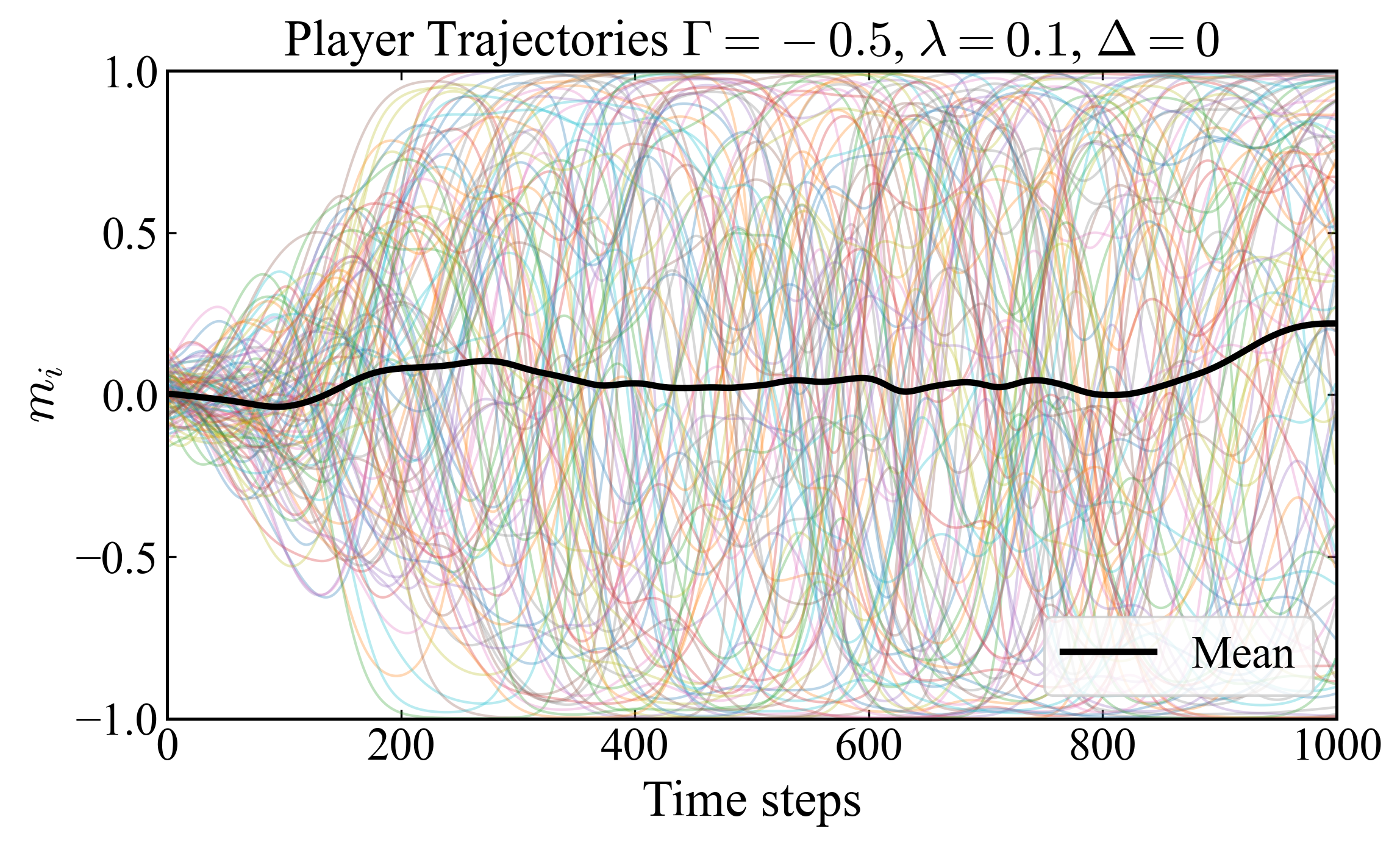}
\caption{Simulated trajectories of a game with no intrinsic bias ($\Delta=0$). Each line represents the time series corresponding to the $m$ of each player. The upper panel is for $\lambda=0.3$ (quick memory loss, stable phase), the lower panel show the same game, but with a learning dynamics with longer memory ($\lambda=0.1$, chaotic phase).  Both games are identical (have the same payoff matrices and parameters), with the exception of $\lambda$.
In the simulations there are $N = 100$ players, $\beta = 0.01 \times \sqrt{N-1} \approx 0.099$. The game is moderately competitive ($\Gamma=-0.5$).
In all figures we use $\sigma=1$.}
\label{fig:combined1}
\end{figure}

To illustrate typical trajectories in the stable phase we focus on one particular instance of a game with no bias towards any action ($\Delta=0$) in Fig.~\ref{fig:combined1}. In the example, there are $N=100$ players and the pairwise games are moderately competitive ($\Gamma=-0.5$). In the upper panel the players are subject to relatively quick memory loss (unique-fixed point phase). The dynamics converges to a state in which all players choose each of their actions at random with probability $1/2$ (i.e., $m_i=0$ for all $i$). This is the paramagnetic state described in \cite{garnier2024unlearnable}. For longer memories (lower panel, unstable phase) the dynamics remains volatile and potentially chaotic.

\begin{figure}[t]
\centering
\includegraphics[width=\linewidth]{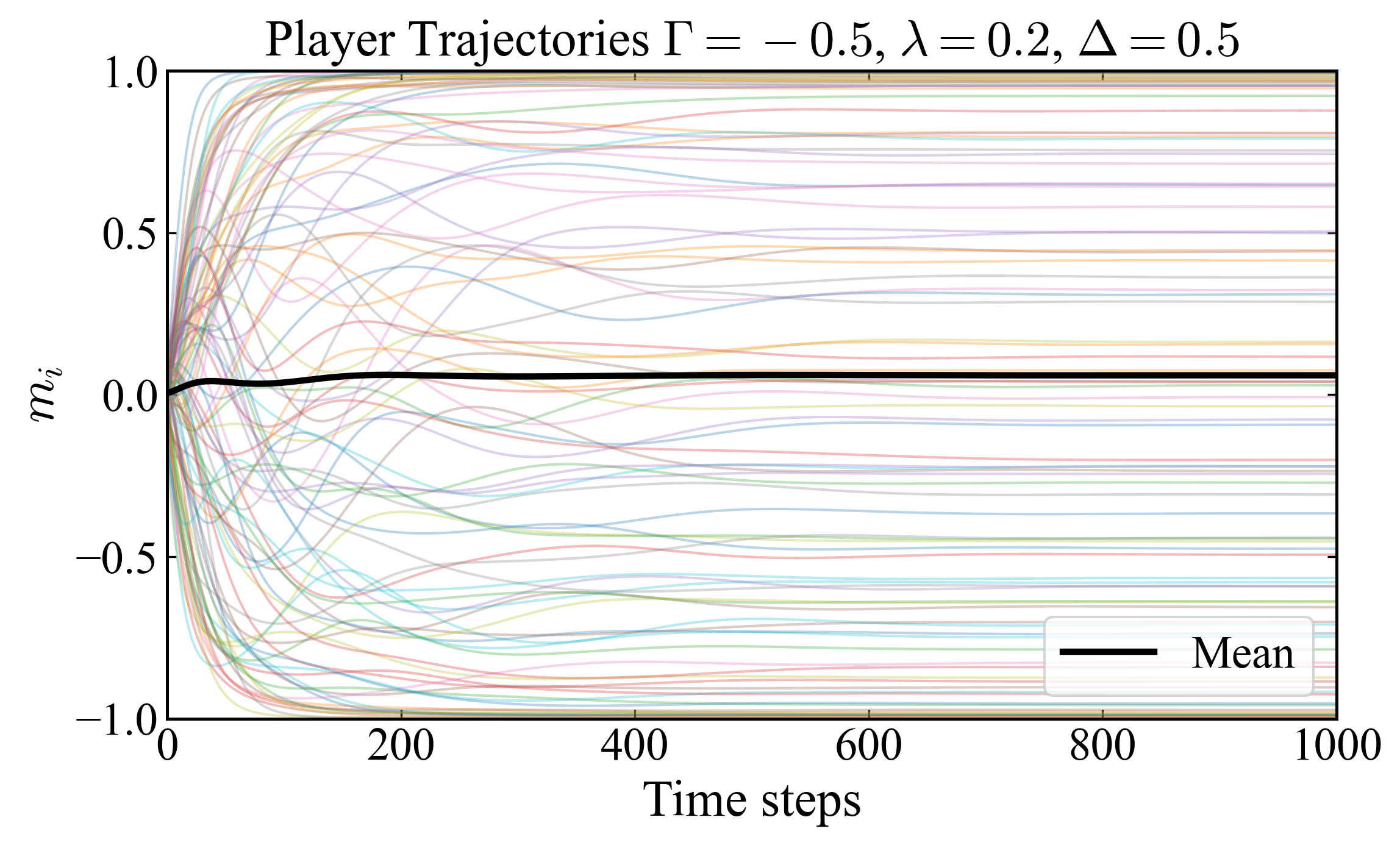}

\vfill
     \includegraphics[width=\linewidth]{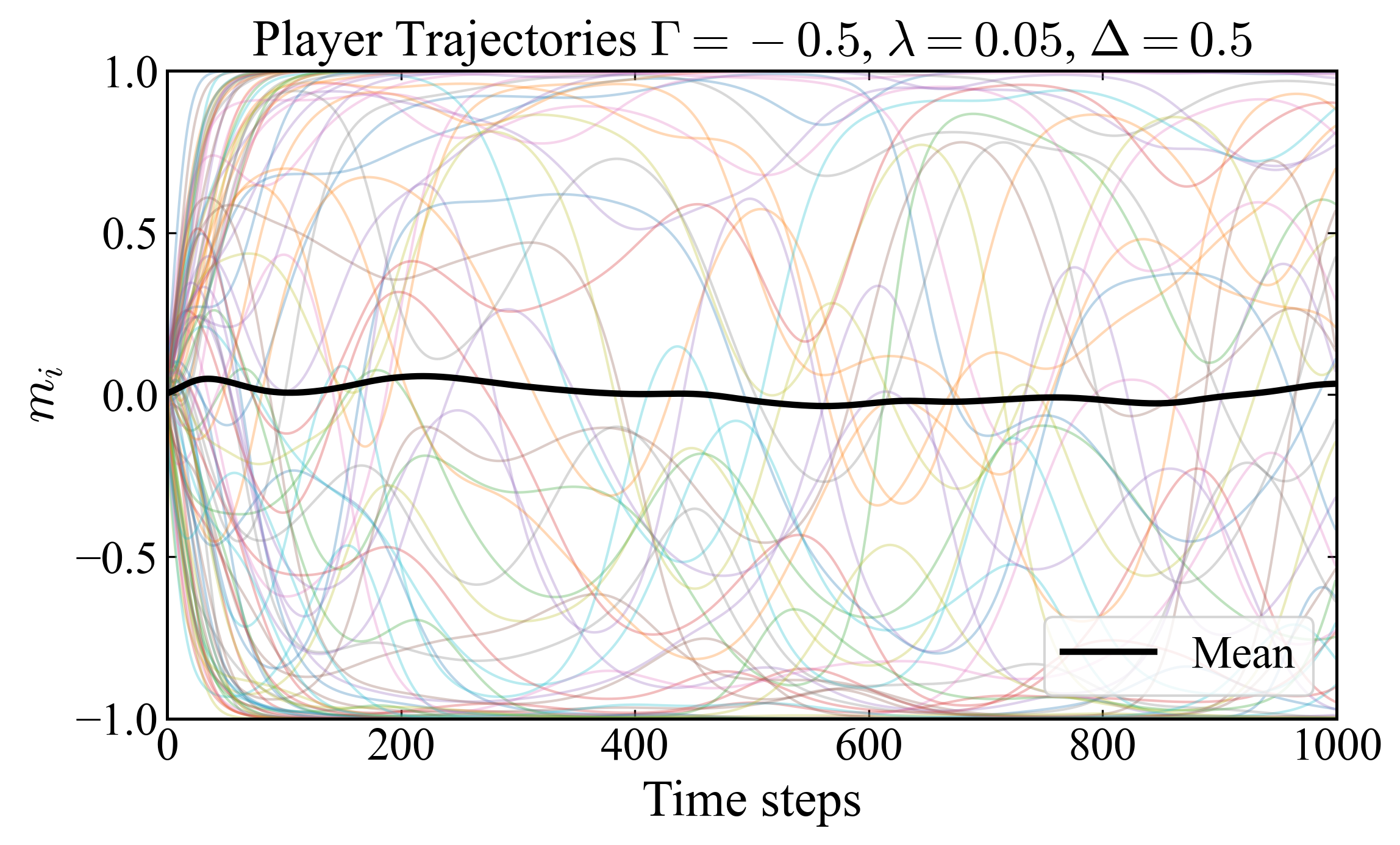}
\caption{Simulated trajectories for a game with non-zero bias ($\Delta=0.5$). Upper panel is for quick memory loss ($\lambda=0.2$, stable phase), lower panel for longer memories ($\lambda=0.05$, chaotic phase). Remaining parameters are as in Fig.~\ref{fig:combined2}.}
\label{fig:combined2}
\end{figure}

Fig.~\ref{fig:combined2} shows similar trajectories but now for a biased game ($\Delta\neq 0$). For long memories (lower panel) the dynamics is again irregular and does not settle down. When $\lambda$ is sufficiently large (memory loss sufficiently quick, upper panel) we find again a phase in which each random game has a unique stable fixed point. In contrast to the unbiased game (Fig.~\ref{fig:combined1}) the distribution of $m_i$ at the fixed point is spread out in the interval $(-1,1)$ though. This means that each player plays one of their actions more often than the other.

\begin{figure}[h]
\centering
\includegraphics[width=0.95\linewidth]{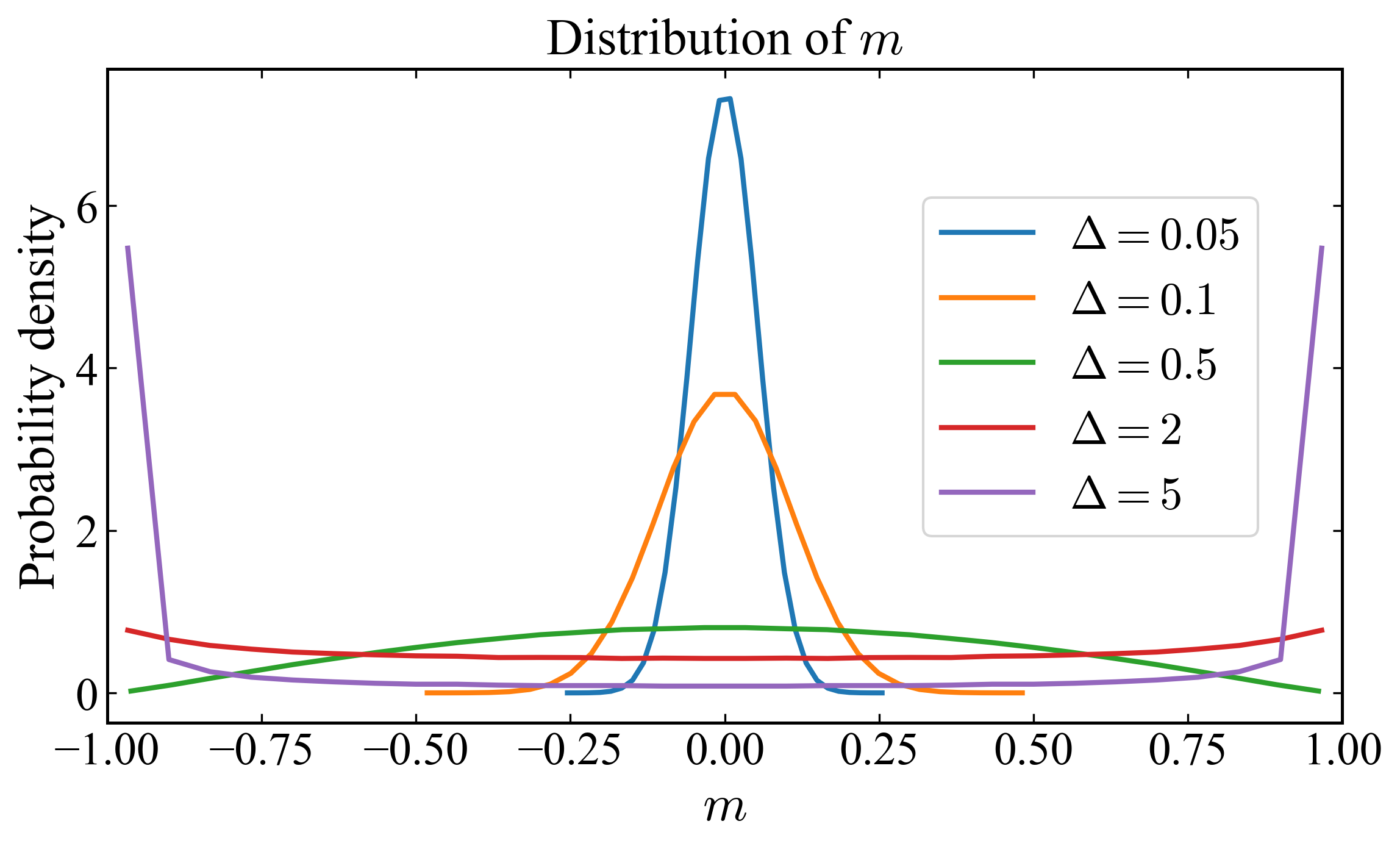}
\caption{Distribution of the $\{m_i\}$ in the unique fixed point phase, for different magnitudes $\Delta$ of the random field. Results are for $\Gamma=-0.5, \lambda=0.5$, and are obtained from the analytical theory by solving the self-consistency relations in the fixed-point phase (see Sec.~\ref{sec:fp}), and then plotting the distribution of the zeroes $m(z)$ of Eq.~(\ref{eq:fp}), with $z$ a standard Gaussian random variable. }
\label{fig:P_of_m}
\end{figure}

The distribution of the $\{m_i\}$ in the fixed-point phase is shown for different values of the bias parameter $\Delta$ in Fig.~\ref{fig:P_of_m}. For small biases the distribution is concentrated on $m=0$. All players then play each of their strategies with more or less the same probability (mixed strategies are near $x_i=1/2$). As the random bias is increased (increasing $\Delta$) the distribution of the $m_i$ across players spreads out, and eventually becomes bimodal. For large biases each $m_i$ is either near $-1$ or near $1$, indicating that each player predominantly plays one of their two actions.


\begin{figure}
\includegraphics[width=0.9\linewidth]{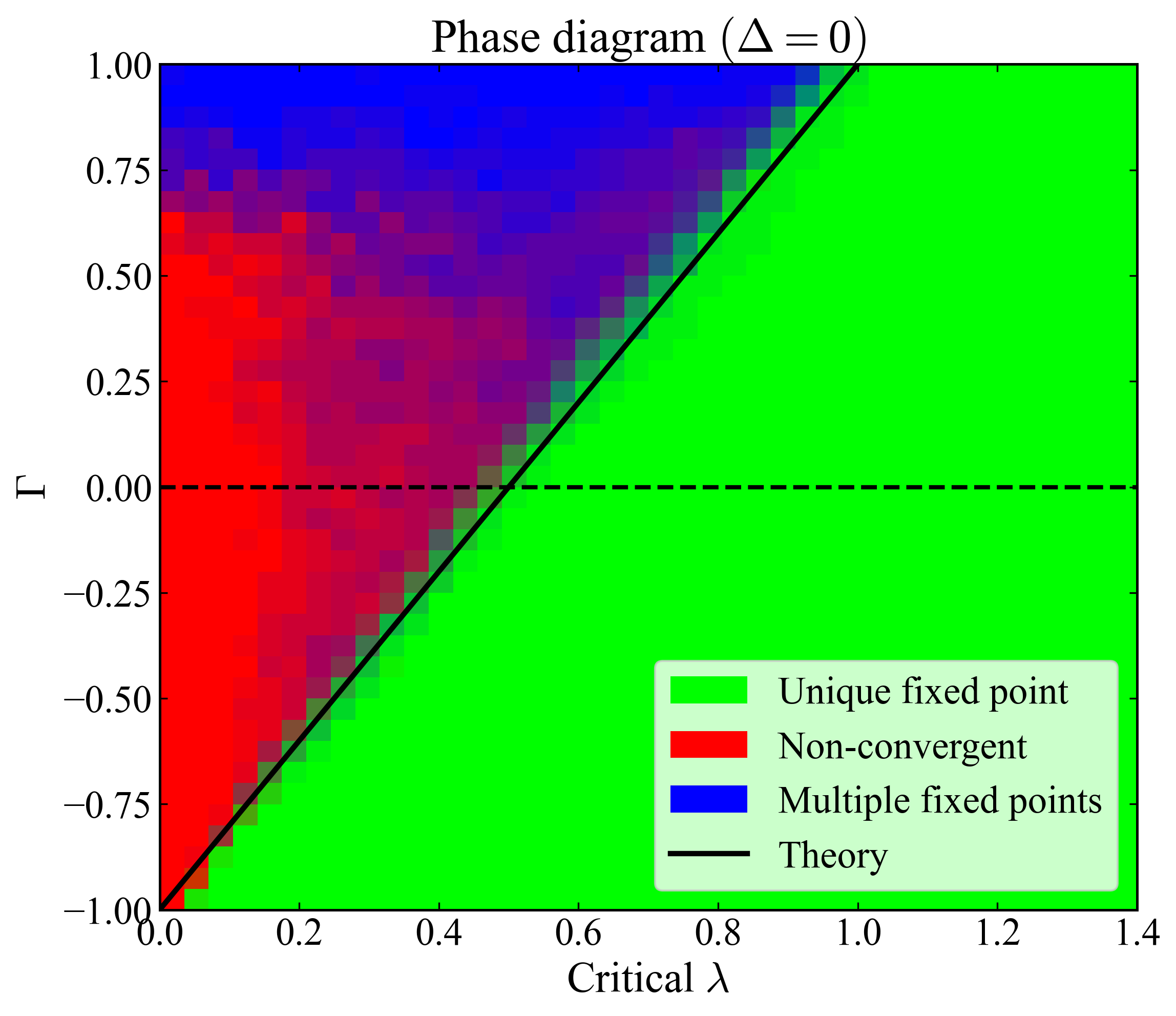}
\includegraphics[width=0.9\linewidth]{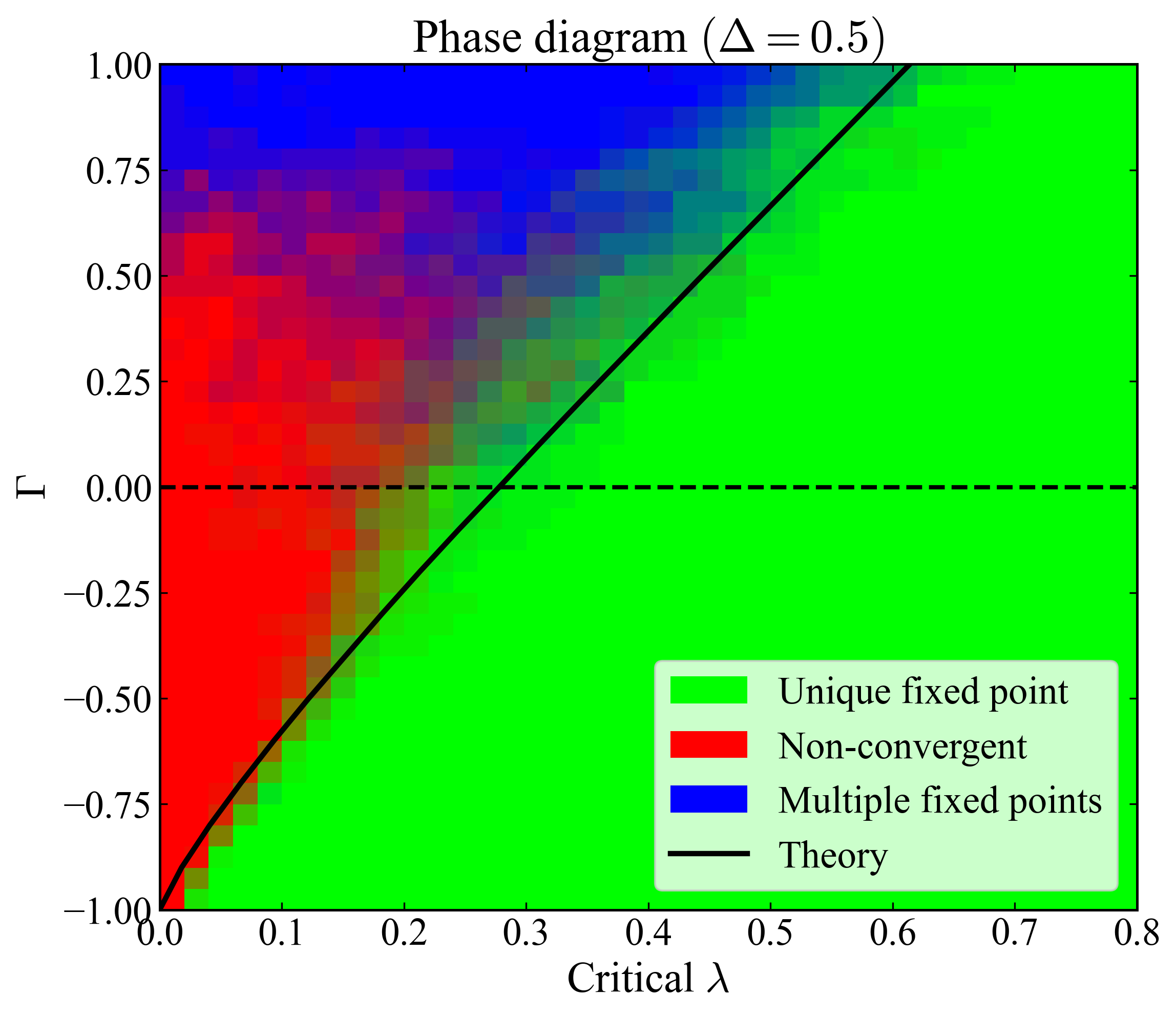}
\caption{Empirical phase diagrams for two different choices of $\Delta$. For each parameter setting, we simulate 20 independent games, each with 40 distinct initial conditions. Each game consists of 100 players with $\beta = 0.01 \times \sqrt{99} \approx 0.099$. We note the different scales on the horizontal axis in the two panels. A game is classified as: (i) a unique fixed point (green), if all trajectories converge to the same fixed point; (ii) non-convergent (red), if no trajectories converge; or (iii) multiple fixed points (blue), if trajectories converge to more than one distinct fixed point.
Each simulation is run for up to 50,000 steps. A trajectory is deemed converged if, over five consecutive steps, the Euclidean distance between successive updates falls below $10^{-6}$. A game is classified as having a unique fixed point if all 20 trajectories converge to the same solution, identified numerically when the pairwise Euclidean distance between fixed points is less than $0.01$.\label{fig:pg_sk}}
\end{figure}

\subsection{Stability diagram and role of the random field}
We now discuss the stability behaviour in more detail. Whether or not the learning dynamics is stable (in the limit of a large number of players) for a particular parameter combination can be determined by first solving the self-consistency relations in Eq.~(\ref{eq:self_consistency}) and then testing the stability condition in (\ref{eq: stab_relation_main}). From this we obtain the stability boundary as a function  of $\Gamma$, $\lambda$ and $\Delta$. 

The resulting stability diagram is shown in Fig.~\ref{fig:pg_sk}. The solid black line is the stability boundary, determined from the theory. This is where fixed points become linearly unstable. For each fixed value of the bias parameter $\Delta$ the learning dynamics is predicted to converge to a unique fixed point to the right of the respective stability line. To the left, we expect either multiple fixed points or persistently volatile dynamics. 

These predictions are confirmed by simulation results, also shown in Fig.~\ref{fig:pg_sk} (background colouring). For each point in parameter space, we compute the fraction of games falling into each category (unique fixed point, multiple fixed points, or persistently volatile behaviour) and assign these fractions to the corresponding colour channels: red for non-convergent behaviour, green for a unique fixed point, and blue for multiple fixed points. The final colour at each point is obtained by combining these three components in an RGB colour scale as was previously done for Lotka--Voltera systems with random interactions in \cite{sidhom}. This leads to a continuous colour representation of the empirical phase behaviour. Pure colours (red, green or blue) indicate regions where one type of behaviour dominates, while mixed colours correspond to coexistence of different dynamical outcomes across disorder realisations. 

We find that the learning dynamics nearly always converges to unique fixed points to the right of the stability lines (when memory loss is sufficiently quick). When players have longer memories ($\lambda$ sufficiently small), there are either many fixed points or the dynamics does not settle down. This persistently volatile behaviour is mostly observed for sufficiently competitive games ($\Gamma\lesssim 0$ and/or memory loss very weak, red region in Fig.~\ref{fig:pg_sk}), whereas multiple fixed points are mostly see for cooperation-type games ($\Gamma >0$, blue region).

\begin{figure}[h]
            \centering
        \includegraphics[width=0.9\linewidth]{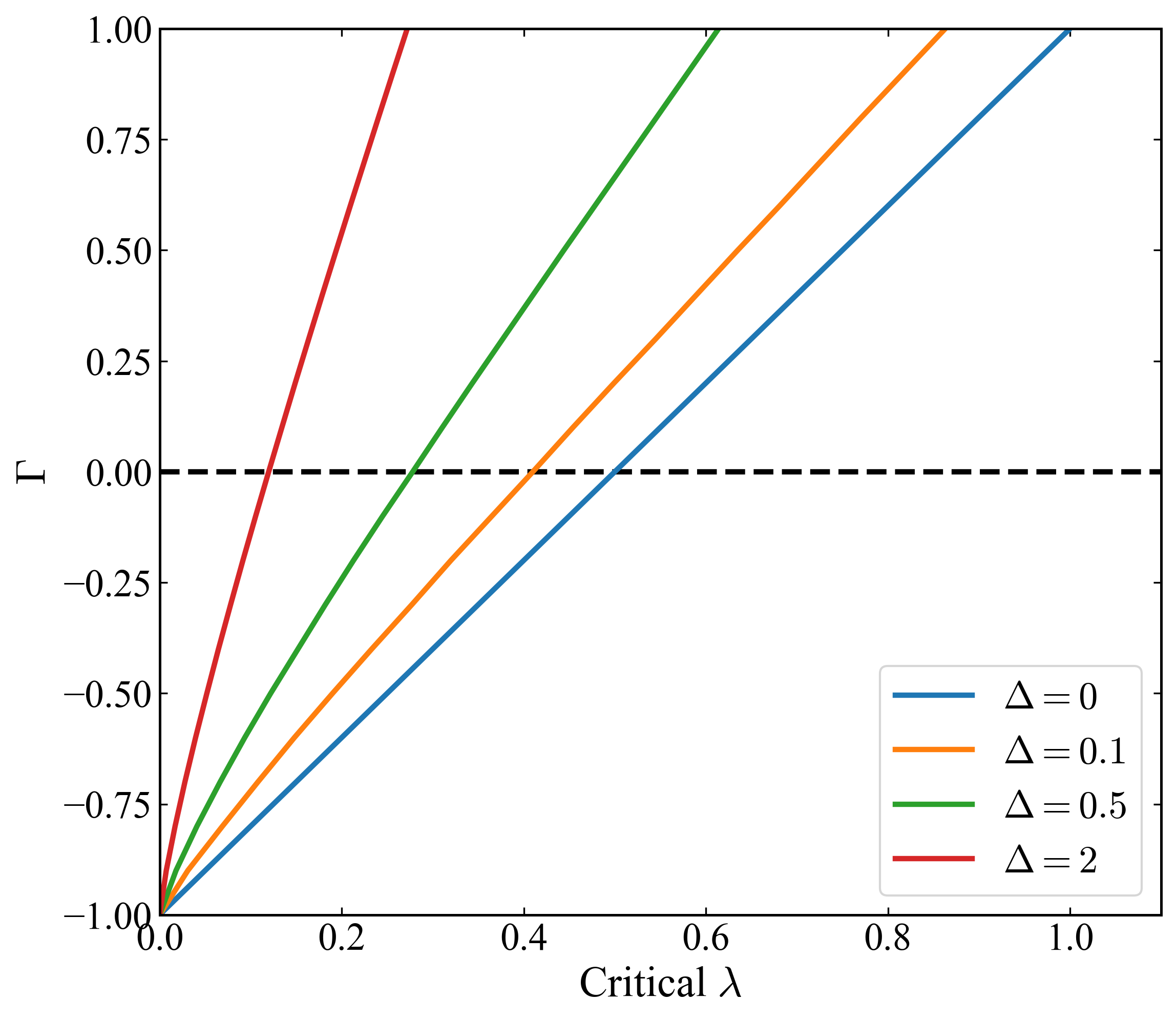}
        \caption{Stability boundary for the SK game with random field, as a function of the memory-loss and competition parameters $\lambda$ and $\Gamma$. The lines indicate the onset of linear instability as obtained from the theory (see Sec.~\ref{sec:lin_stab_main}), for different magnitudes $\Delta$ of the random field. As can be seen from the figure, the stable region (to the right of the respective lines) becomes larger with increasing field amplitude.}
        \label{fig:stab_diagram_delta}
    \end{figure}

Interestingly increasing the bias parameter $\Delta$ makes the learning more stable (the stability line is moved to the left as shown in Fig.~\ref{fig:stab_diagram_delta}, i.e., the stable region becomes larger). A strong bias parameter effectively anchors each player toward a preferred action, which makes volatile motion or the existence of multiple fixed points less likely. This promotes stability. 

Mathematically, the stabilising effect can also be understood as follows. Introducing a random field (i.e., choosing $\Delta > 0$), `smears out' the distribution of the $m_i$ at the fixed point (see Fig.~\ref{fig:P_of_m}), and players become increasingly `frozen' (i.e., they tend to use only one of their actions, $m_i$ close to $+1$ or $-1$). The stability condition in Eq.~(\ref{eq: stab_relation_main}) can be written as (see Appendix~\ref{sec:lin_stab_app})
\begin{align}\label{eq: stab_relation_2}
      \sigma^2 \left\langle \left[ \frac{(1 - m^2) } { 2\lambda - (1 - m^2) \Gamma\sigma^2 \chi } \right]^2 \right \rangle < 1
\end{align}
As the fixed-point values $m$ move towards $\pm1$, the numerator $1-m^2$ tends to zero, whereas the denominator in the average remains finite (for $\lambda>0$). Thus, the expression on the left-hand side of (\ref{eq: stab_relation_2}) reduces, making the system more stable.

\section{Grand-canonical SK game}\label{sec:gcsk}
\subsection{Definition}
We now take inspiration from the (grand-canonical) Minority Game (MG) \cite{challet2006minority}, and define the grand-canonical SK game. In this game there are two types of agents. Using the terminology from the Minority Game literature \cite{challet2000modeling} these are `speculators' and `producers', respectively. Each speculator only has one active strategy and must decide, at each step, whether they want to play or not. If a speculator plays, they can only use that single strategy. The only other option is to abstain. A speculator performs an adaptive learning process similar to what we described above. They play if they expect their payoff to exceed a certain threshold (to be defined later in detail). Producers also only have one single strategy, but they must play no matter what, i.e., they do not have the option of abstaining.

We assume that the total number of players is $N$. The number of speculators is $N_s = s N$ ($0\leq s\leq 1$), and the remaining $N_p=(1-s)N$ agents are producers. Without loss of generality we assume that the speculators have labels $i=1,\dots,N_s$, and the producers are the agents $i=N_s+1,\dots, N$. 

We assume that the payoff to agent $i$ is $a_{ij}$ if they interact in a game with player $j$. For this interaction to occur, both players need to actually play in that iteration. We write $x_i(t)$ for the probability that player $i$ plays in round $t$. This quantity can be time-dependent (due to the learning dynamics) for $i=1,\dots, N_s$ (the speculators). The producers $i=N_s+1, \dots, N$ always have $x_i=1$.

Each speculator has a propensity to play, $q_i(t)$. This propensity determines the probability with which the speculator plays, via
 \be
x_i(t)=\frac{e^{\beta q_i(t)}}{1+e^{\beta q_i(t)}}.
\ee
In the limit $\beta\to\infty$ player $i$ plays with probability one if $q_i>0$, and with probability zero if $q_i<0$.

The propensities are updated as follows,
\BE\label{eq:gc_update}
q_i(t+1)&=&(1-\alpha) q_i(t)+\Pi_i[\bx(t)]-\varepsilon,
\EE
where $\alpha$ is again memory loss, and where the payoff player $i$ can expect if they decide to play (and given everybody else's mixed strategies) is given by
\be
\Pi_i(\bx)=\sum_{j=1,\dots, N_s; j\neq i}a_{ij} x_j(t)+ \sum_{j=N_s+1}^N a_{ij}.
\ee
The first sum comes from games with the other speculators, and the second sum is from games with the producers. 

The quantity $\varepsilon$ can be interpreted as a `cost to play'. I.e., participating in the game comes with some cost to the player (e.g., a transaction or processing cost, or participation fee), that is to be subtracted from the payoff. The quantity $\varepsilon$ therefore describes a minimum (expected) payoff speculators require to decide to play the game (setting memory loss aside, the increment of the propensity is positive if $\Pi_i>\varepsilon$, and negative if $\Pi_i<\varepsilon$).

We assume the following statistics for the payoff matrix elements, 
\be
\overline{a_{ij}}=0, ~~~\overline{a_{ij}^2}=\frac{\sigma ^2}{N-1}, ~~~ \overline{a_{ij}a_{ji}}=\frac{\Gamma\sigma^2}{N-1},
\ee
similar to the two-strategy game. 

Taking the continuous-time limit as before and again writing $\lambda=\alpha/\beta$, we find
\BE
\dot x_i &=& x_i(1-x_i)\bigg[-\lambda  \ln \frac{x_i}{1-x_i}+ \sum_{j=1}^{N_s} a_{ij} x_j\nonumber \\
&&~~~~~~~~~~~~~~~~~+ \sum_{j=N_s+1}^N a_{ij}-\varepsilon\bigg].
\EE
The term from the interaction with the producers acts as an external field $
h_i = \sum_{j=N_s+1}^N a_{ij}$, and we have $\overline{h_i^2} = (1-s)\sigma^2$. We also have $\overline{h_i h_j}=\Gamma\sigma^2/(N-1)$ , but this is sub-leading and does not contribute to the dynamics in the limit $N\to\infty$.
\subsection{Generating-functional analysis}
Carrying out a standard generating-functional calculation (see Appendix \ref{app:gcsk_gfa}) we find the dynamic mean field process:
\BE
\dot x &=& x (1-x) \bigg[-\lambda \ln\frac{x}{1-x}+s\Gamma\sigma^2\int dt' G(t,t') x(t')\nonumber \\
&&~~~~~~~~~~~-\varepsilon+\eta(t)\bigg],
\EE
with
\be \label{eq:self_con_canon}
\avg{\eta(t)\eta(t')}= \sigma^2 \left[s \avg{x(t)x(t')}+(1-s)\right],
\ee
and with the response function
\be
G(t,t')=\avg{\frac{\delta x(t)}{\delta \eta(t')}}.
\ee
The notation $\avg{\cdots}$ again stands for an average over realisations of the effective process (i.e., over realisatons of noise trajectories $\{\eta(t)\}$).

The model parameters are $\lambda, s, \Gamma, \sigma^2$ and $\varepsilon$, but fixed points and their stability only depend on $\lambda/\sigma$, $\varepsilon/\sigma$, $\Gamma$ and $s$.  Through all figures for this model we always use $\sigma = 1 $.

We conduct a similar analysis as in the previous sections, first by assuming the dynamics have a stable fixed point; $x$ and $\eta$ then become static random variables. This time setting $\eta = \sigma\sqrt{sq + (1-s)}\,z$, we obtain
 \be\label{eq:cont_gc}
-\lambda \ln \frac{x}{1-x} + s\Gamma\sigma^2 \chi x  - \varepsilon + \sigma   \sqrt{qs + (1-s)} z=0.
 \ee

The  corresponding self-consistency equations are given by [with $x(z)$ the solution of Eq.~(\ref{eq:cont_gc})]
\BE\label{eq:sc_gcsk}
q &=& \int Dz\, x(z)^2, \nonumber \\
\chi &=& \frac{1}{\sigma \sqrt{qs + (1-s)} } \int Dz\, \frac{\partial x(z)}{\partial z},
\EE
where we write again $Dz = \frac{dz}{\sqrt{2\pi}} e^{-z^2/2}$.

 Conducting a similar linear stability analysis as in the previous section, we find the fixed point is linearly stable when
 \be\label{eq:stab_gc}
      \sigma^2 s \left\langle \left(\frac{\partial x } { \partial \eta} \right) ^2 \right \rangle < 1
  \ee
A full derivation can be found in Appendix~\ref{app:gcsk_gfa}.

\subsection{Stability diagram and dynamical behaviour}
\begin{figure}
\centering
     \includegraphics[width=0.9\linewidth]{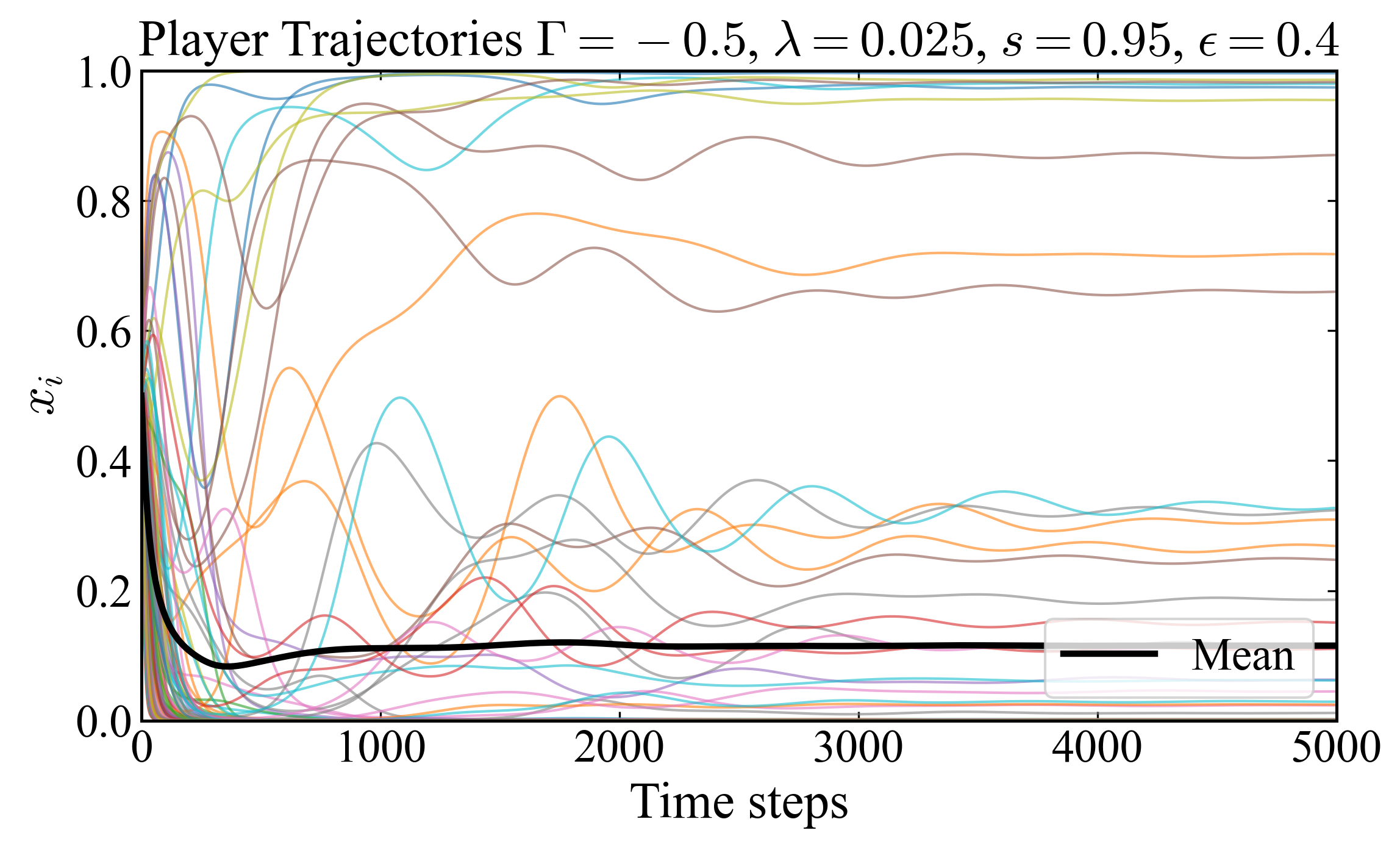}
\vfill
     \includegraphics[width=0.9\linewidth]{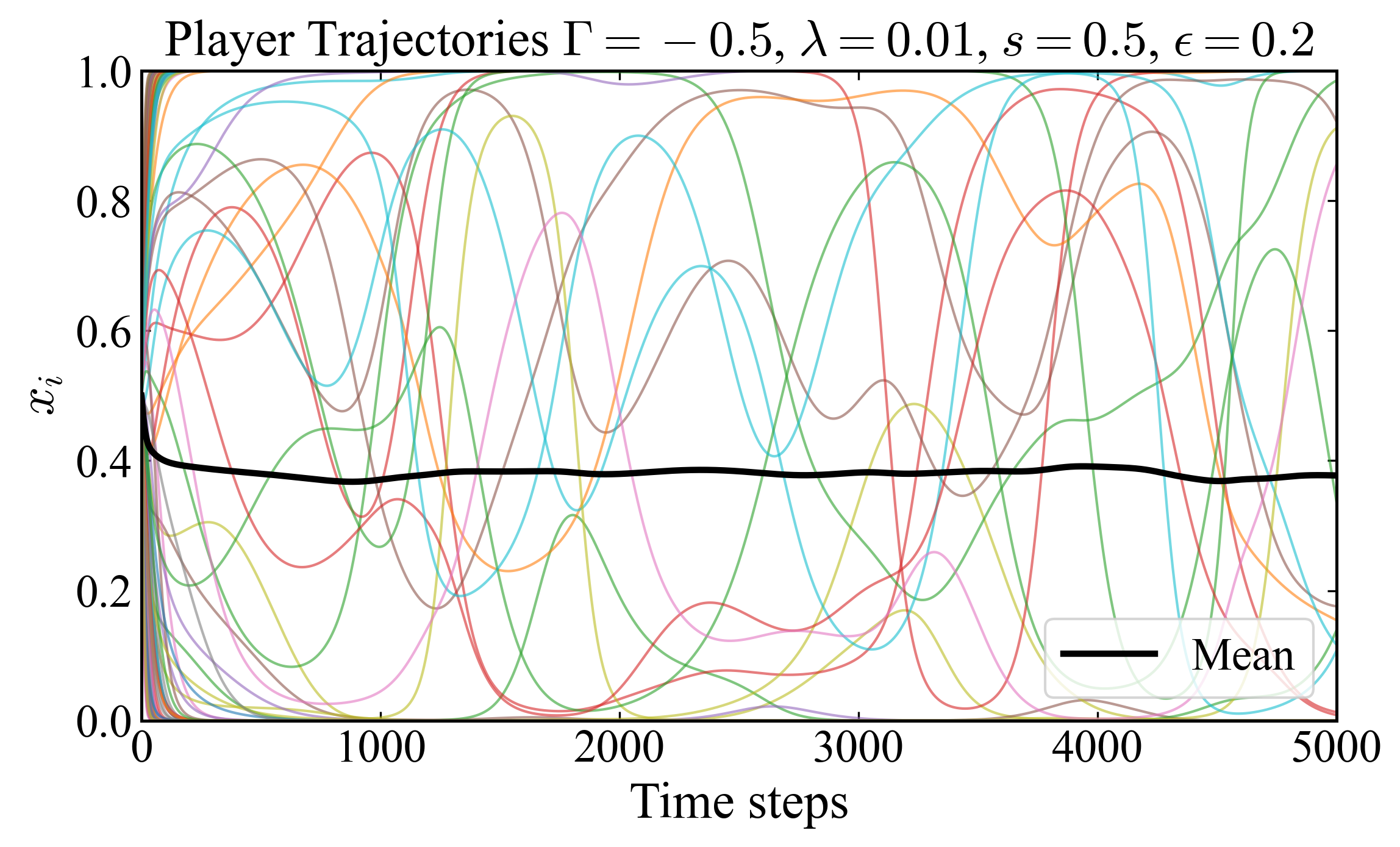}
\caption{Simulated trajectories of two different grand canonical games with different parameters with one converging to the unique fixed point and the other exhibiting chaotic dynamics.  Each simulation consists of $100$ speculators and each line represents the time series corresponding to the likelihood each speculator will do an action the given time. ($\beta = 0.1$, $\sigma = 1 $).}
\label{fig:combined5}
\end{figure}

Similar to the previous setup, we find that the grand-canonical SK game has three dynamical regimes: (i) a phase with a unique stable fixed point for fixed instances of the game, (ii) a volatile regime, and (iii) a regime in which each game has many fixed points.  Although the critical $\lambda$ is different and depends on additional parameters, the qualitative picture in terms of $\Gamma$ and $\lambda$ for any given choice of parameters remains similar to that shown in Fig.~\ref{fig:phase_diag}. Figure \ref{fig:combined5} illustrates two instances of learning in the the grand-canonical SK game: one in which the dynamics converges, and another in which it exhibits volatile dynamics. 

\begin{figure}[t]
\centering
     \includegraphics[width=\linewidth]{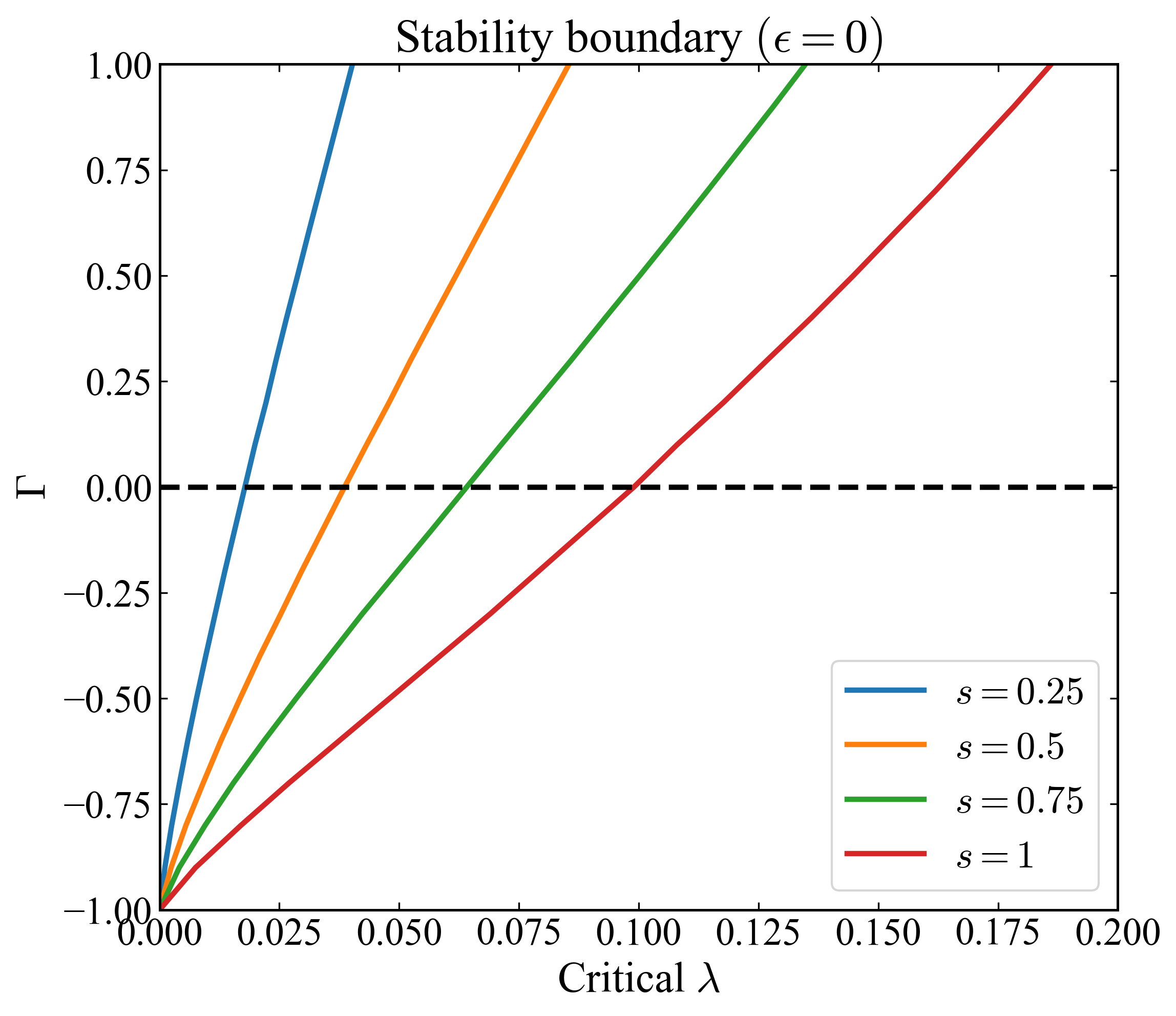}
\vfill
\includegraphics[width=\linewidth] {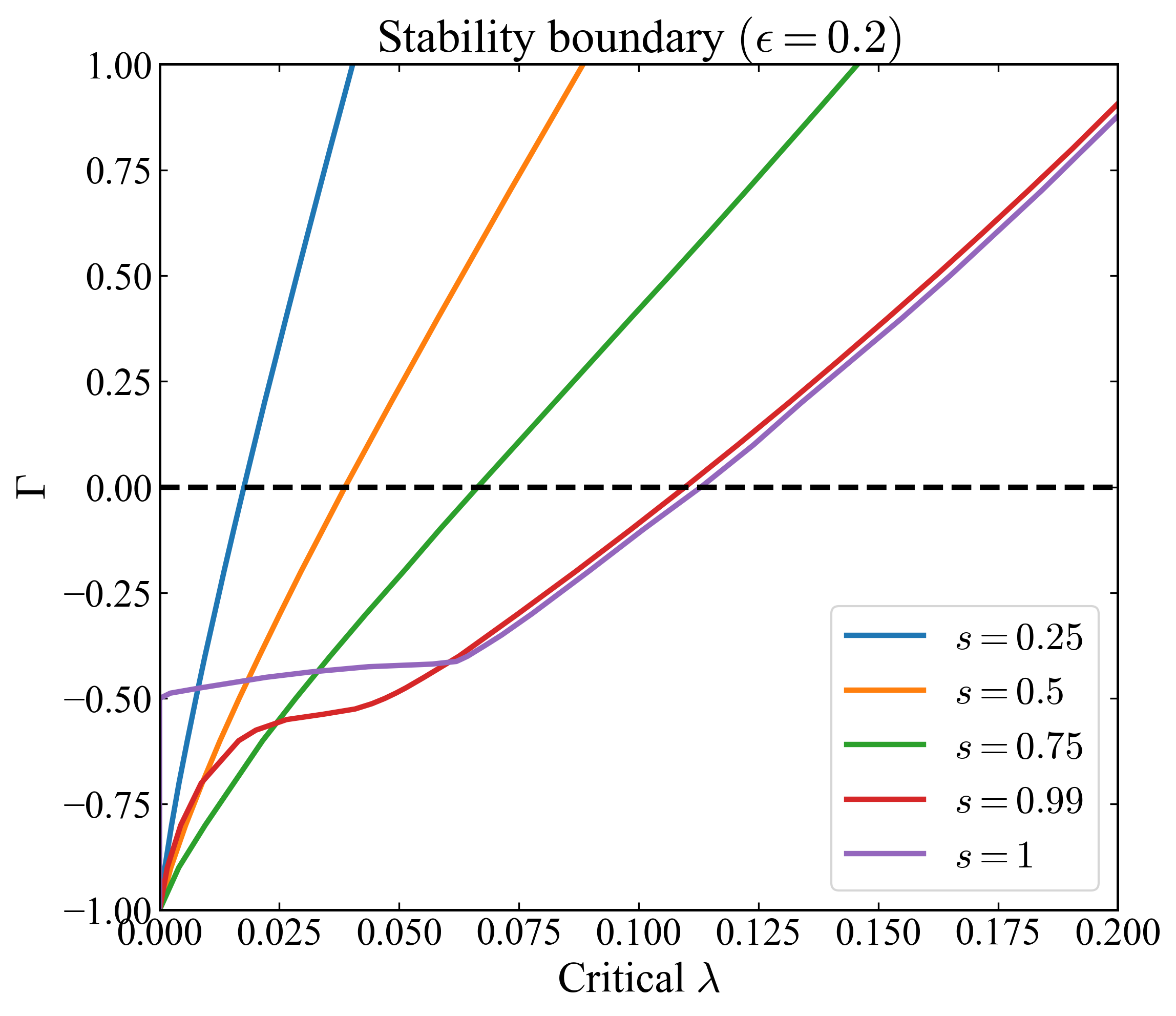}
\caption{Stability boundary of the grand canonical SK game in the $(\lambda,\Gamma)$ plane (the memory loss and competition parameters, respectively), as predicted from the theory for varying $s$ for $\varepsilon = 0$ (upper panel) and $\varepsilon = 0.2$ (lower panel). The learning dynamics converges to unique fixed points to the right of the relevant line. To the right we find either chaotic dynamics or the learning process has many stable fixed points. }\label{fig:stab_gcsk1}
\end{figure}

Similar to the previous sections, we can determine the stability diagram by solving the self-consistency equations (\ref{eq:sc_gcsk}) and then testing the stability condion in~(\ref{eq:stab_gc}). Results in the plane spanned by the memory loss parameter $\lambda$ and the competition parameter $\Gamma$ are shown in Fig.~\ref{fig:stab_gcsk1} for different values of the payoff threshold $\varepsilon$, and different fractions $s$ of speculators in the population of players.

We note, when $\varepsilon = 0$ (upper panel of Fig.~\ref{fig:stab_gcsk1}), the phase diagram for the grand-canonical SK game takes a similar shape to the SK game with a random field.  We find again stability for sufficiently quick memory loss (sufficiently large $\lambda$). Making the game more competitive (reducing $\Gamma$) makes learning more stable similar to what was found for complex two-player games in \cite{galla2013complex}. We also find that a reduction of the proportion of speculators $s$ leads to a larger stable region. This is consistent with the first relation in Eq.~(\ref{eq:sc_field}) and Eq.~(\ref{eq:self_con_canon}) from which one can see that $1-s$ is analogous to the field amplitude $\Delta$. An increase of the fraction of producers is similar to an increase of the external field, and stabilises the dynamics.

\begin{figure}
\includegraphics[width=\linewidth]{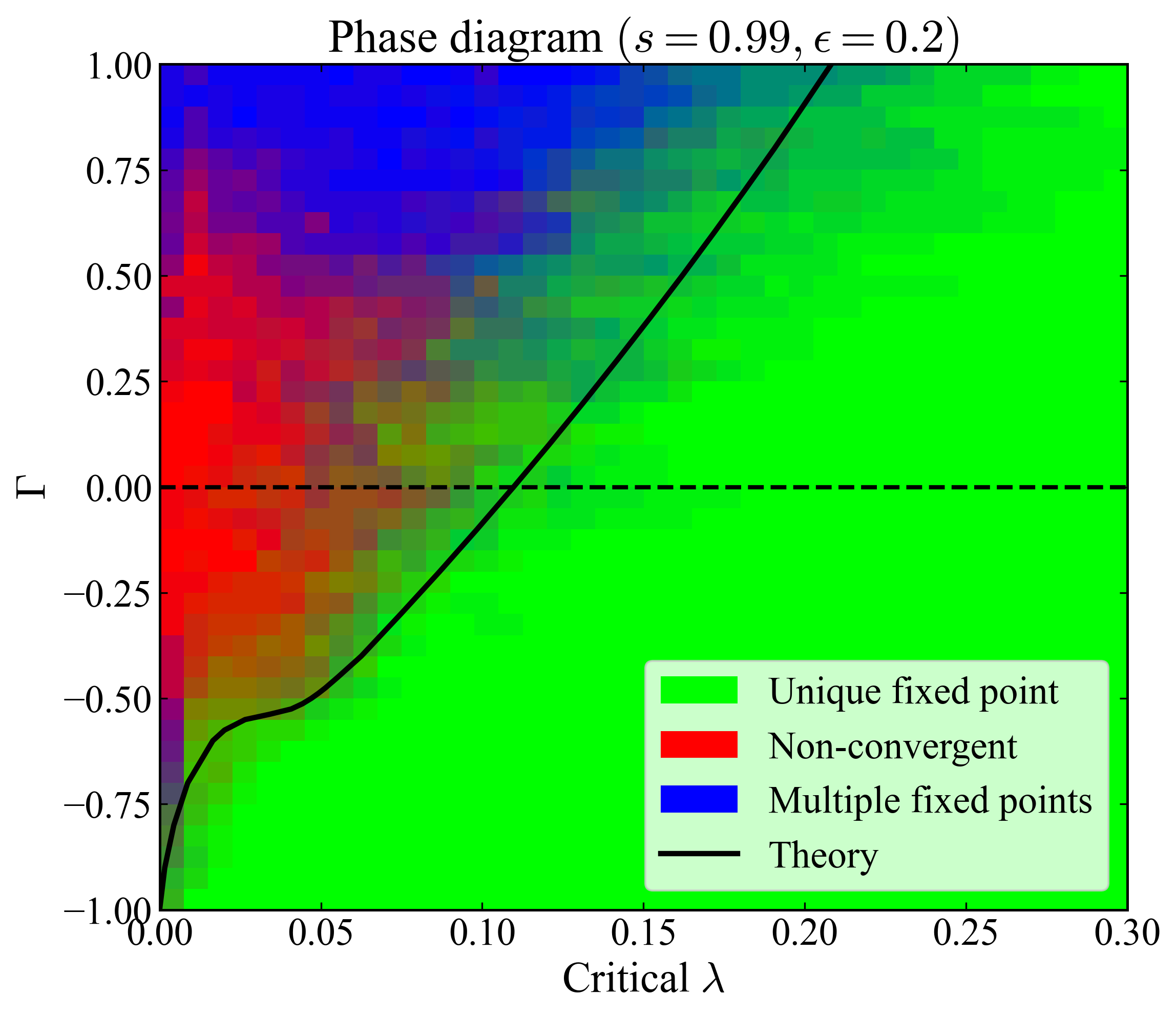}
    \centering
    \vfill
\includegraphics[width=\linewidth]{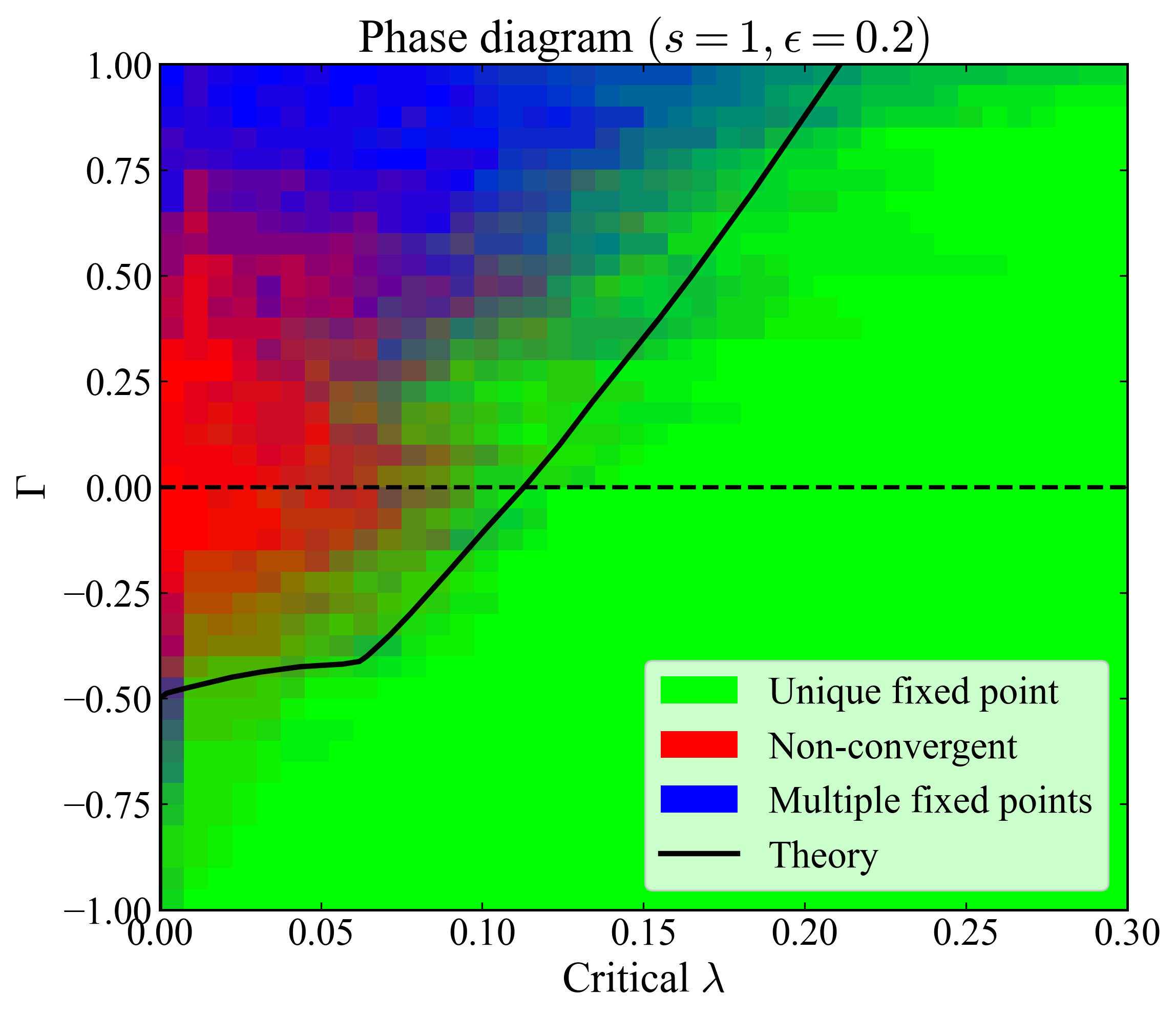}
    \caption{Empirical phase diagrams for two different choices of $s$ and $\varepsilon$. For each parameter setting, we simulate 20 independent games, each with 40 distinct initial conditions. Each game consists of 100 players with $\beta = 0.01 \times \sqrt{99} \approx 0.099$. A game is classified as: (i) a unique fixed point (green), if all trajectories converge to the same fixed point; (ii) non-convergent (red), if no trajectories converge; or (iii) multiple fixed points (blue), if trajectories converge to more than one distinct fixed point.  We classify the dynamical regimes using the same approach as in Fig.~\ref{fig:pg_sk}}
\label{fig:selected_phase_diagrams}
\end{figure}

When $\varepsilon > 0$ (lower panel of Fig.~\ref{fig:stab_gcsk1}), we find that the shape of the stability boundary changes near $s = 1$. A `kink' emerges in the stability curve for negative $\Gamma$. 
For values of $\Gamma $ near $-1$, the system remains stable for very small $\lambda$, but as $\Gamma$ increases, there is a point where there is a sharp increase in the critical $\lambda$.  For larger $\Gamma$, the relation between $\Gamma$ and $\lambda$ returns to being approximately linear. A consequence of this kink is that for some choices of $\Gamma$ and $\varepsilon$, increasing the fraction of speculators $s$ can help stabilise the game. We will discuss this in more detail in the next section. We note that the kink does not appear to be a numerical artifact, as we verify this behavior with simulations in  Fig.~\ref{fig:selected_phase_diagrams}.


\begin{figure}
\includegraphics[width=\linewidth]{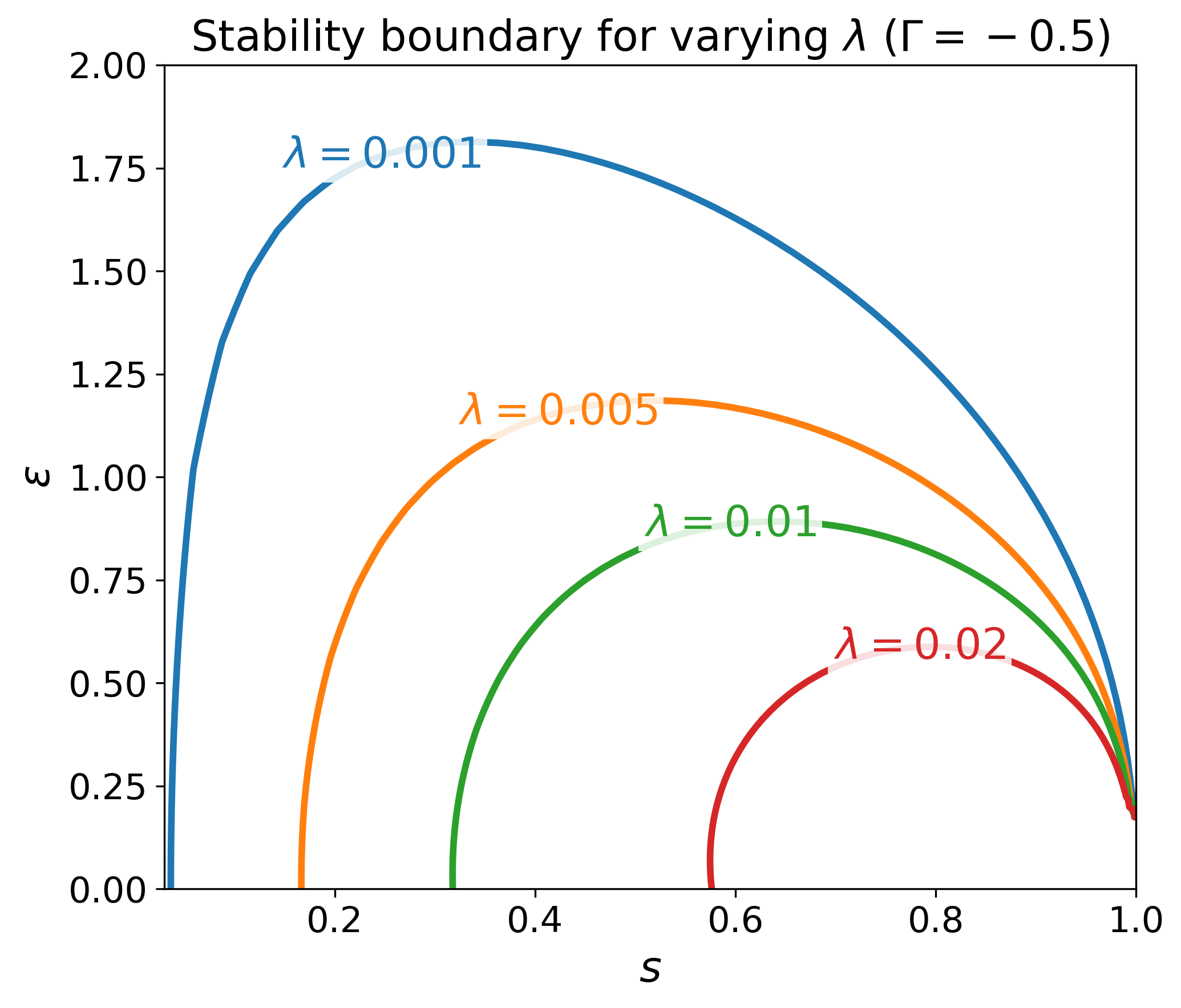}
\caption{Stability boundary for the grand-canonical SK game of varying $\lambda$  for $\Gamma =-0.5$ as a function of $s$ and $\varepsilon$. The stability line is obtained from the analytical theory. The dynamics is predicted to be stable above the stability line.} \label{fig:gcsk_stab_diagr}
\end{figure}
\begin{figure}[h]
    \centering
        \includegraphics[width=\linewidth]{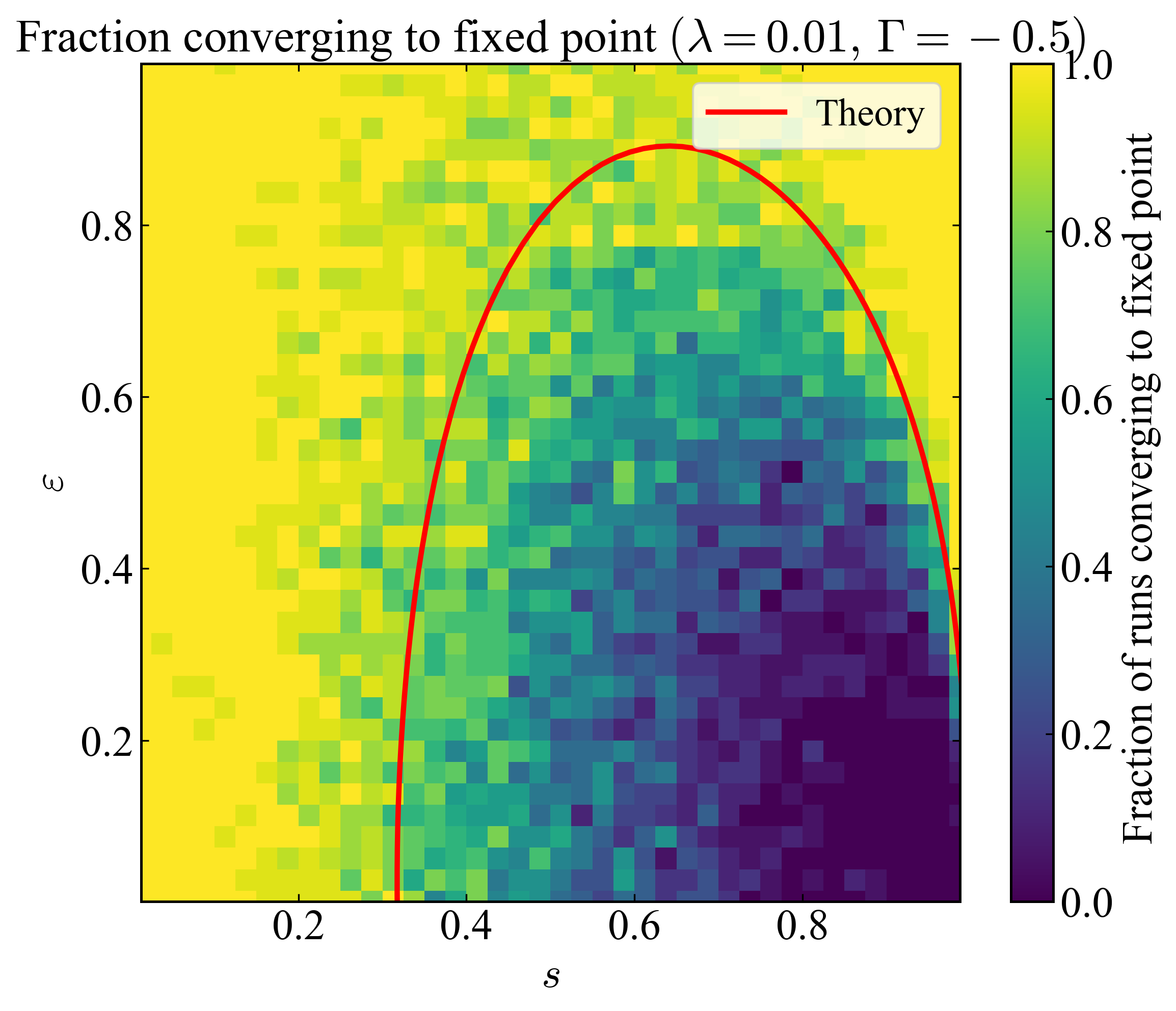}
        \caption{Numerical confirmation of the stability diagram of the grand-canonical SK game, and of the re-entry behaviour as a function of the fraction of speculators $s$. The stability line from the theory is shown in red, learning is predicted to converge above the line. The heatmap shows the number of samples that converged in numerical simulations. For each data point we ran 20 random games for 50,000 steps ($N=100$, $\beta = 0.01 \times \sqrt{99} \approx 0.099, \Gamma=-0.5, \lambda=0.01$)  and recorded how often the learning dynamics converged to a fixed point.}
        \label{fig:phase_diag_s_e+1}
    \end{figure}  
\subsection{Role of the fraction $s$ of speculators, and the payoff threshold $\varepsilon$ on stability}
An interesting picture emerges if we plot the stability diagram in the $(s,\varepsilon)$ plane at fixed $\lambda$ and $\Gamma$. We recall that $s$ is the fraction of speculators among the players, and $\varepsilon$ the threshold payoff speculators expect in order to decide to participate in the games. We show such a stability plot in Fig.~\ref{fig:gcsk_stab_diagr}, for different values of the memory-loss parameter $\lambda$.  The learning process is predicted to be stable (converge to a unique fixed point for any fixed random game) above the stability line. This means that, at a fixed fraction of speculators $s$, we find stability for sufficiently large values of $\varepsilon$. Increasing the payoff threshold $\varepsilon$ reduces how often each speculator participates in the game. This reduced activity in turn stabilizes the learning dynamics.

As can also be seen from Fig.~\ref{fig:gcsk_stab_diagr}, increasing the fraction of speculators $s$ can have a non-monotonic effect on stability. Starting from a small fraction of speculators, adding more speculators tends to destabilise the system. However, once $s$ becomes sufficiently large, a further increase can instead stabilise the learning process. This causes a `re-entry' behaviour in Fig.~\ref{fig:gcsk_stab_diagr}. If we fix a value of $\varepsilon$ that is not too large and start at low $s$ we are in the stable phase. Moving horizontally (constant $\varepsilon$) and to the right in we then enter the unstable phase. Continuing to move to the right (i.e., increasing the fraction of speculators further), we eventually return to the stable phase. This is a reflection of the kink in Fig.~\ref{fig:stab_gcsk1} for $s$ close to one.

We tested these predictions in simulations, results are shown in Fig.~\ref{fig:phase_diag_s_e+1}. The heatmap shows the fraction of samples for which learning converges. The data confirms the re-entry behaviour and is broadly in line with the stability line from the theory. Above the line learning converges for most samples, below the line convergence is rare.

\begin{figure}
        \centering
        \includegraphics[width=\linewidth]{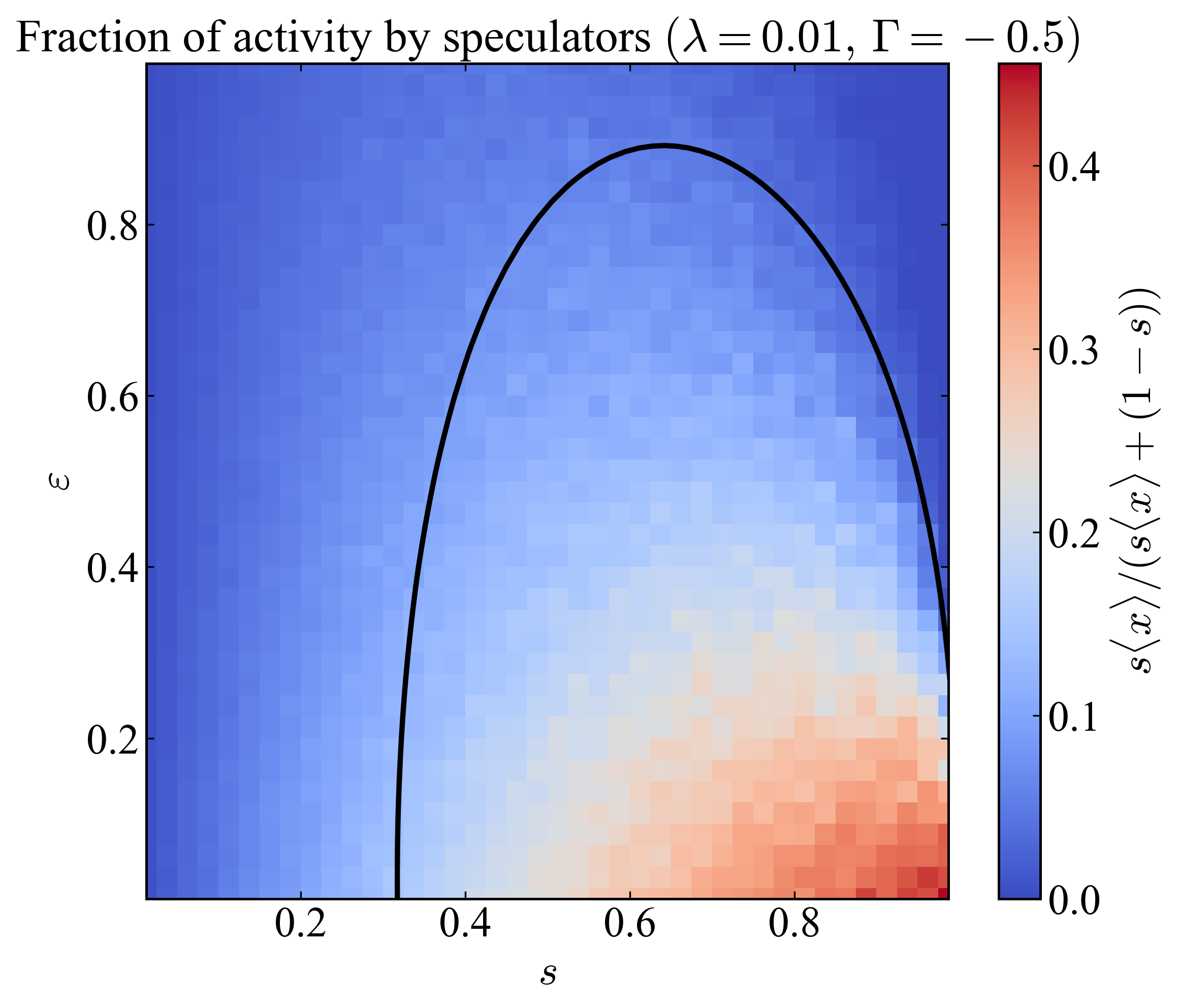}
        \centering
        \includegraphics[width=\linewidth]{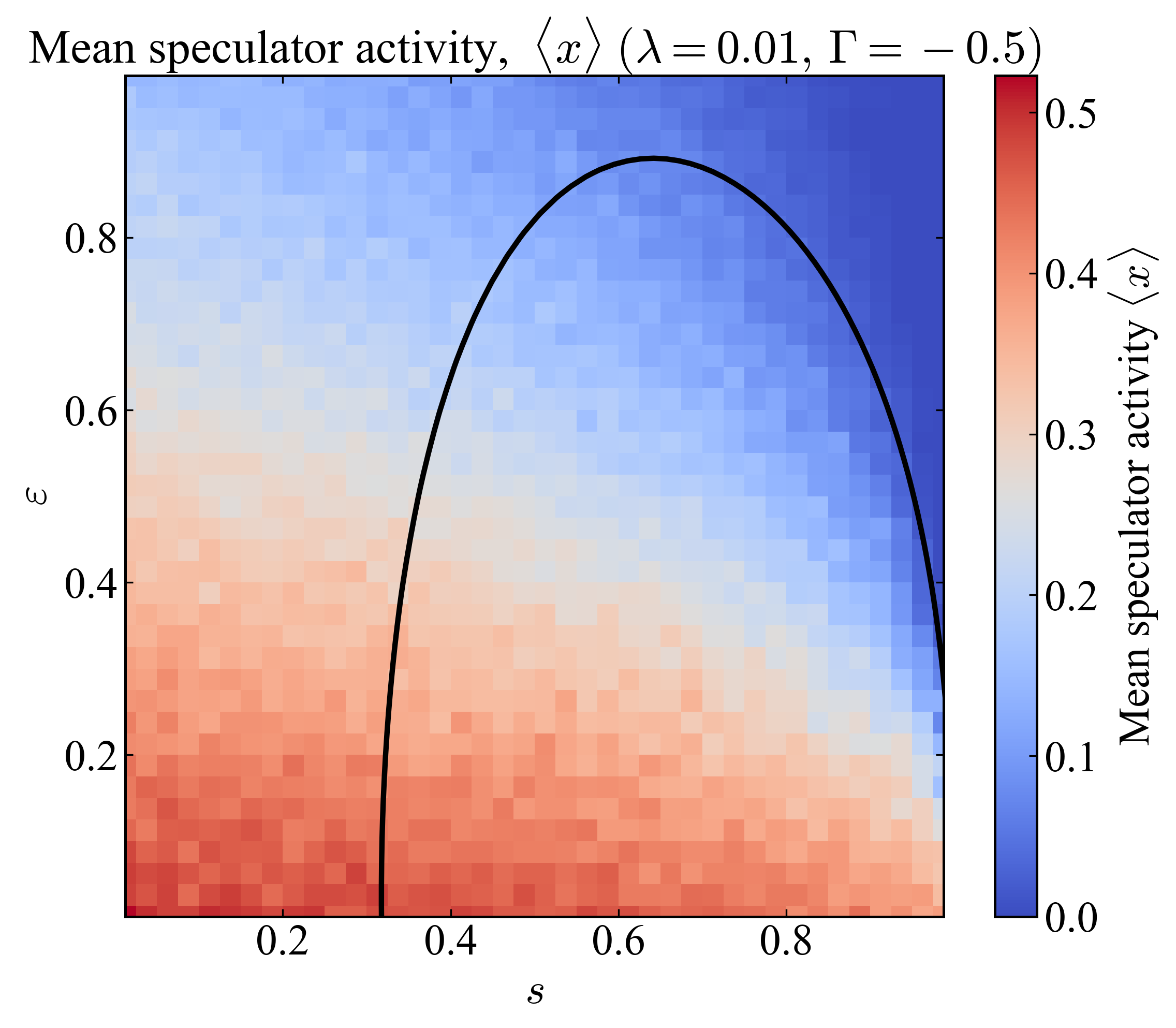}
        \caption{Plot of the fraction of total activity by speculators (upper panel) and mean speculator activity (lower panel) as a function of $s$ and $\varepsilon$. 
        The stability line from the theory is now shown in black.
        The quantities for each grid point are obtained via numerical simulation averaged of 20 games, using the the same parameter choices as Fig. \ref{fig:phase_diag_s_e+1}. The simulations run for 20,000 steps with measurements taken from 10,000 steps onwards.\label{fig:spec_act}}
\end{figure}

To understand this further, we have measured the speculator activity. Results are shown in the upper panel of Fig.~\ref{fig:spec_act}, where we plot the fraction of activity from the speculators relative to the total activity. This is the quantity $s\avg{x}/[s\avg{x}+(1-s)]$, where $\avg{x}$ is the mean probability to play among speculators.  We find that this quantity is relatively low when learning is stable (above the stability line in the figure). A larger proportion of speculator activity appears to be associated with instability. In the lower panel we show $\avg{x}$ for comparison. Naturally, this quantity decreases with an increasing threshold payoff $\varepsilon$. An increased proportion of speculators in the population of players also seems to reduce the mean activity of each speculator. One possible explanation for this latter observation is that increasing the fraction of speculators means that there is a lesser background of producers from which the speculators could profit. 

Mathematically, the re-entrance behaviour as a function of $s$ is driven by the competition between two contributions in the stability condition in (\ref{eq:stab_gc}). We write $F(s)=\sigma^2 s \left\langle \left( \partial x / \partial \eta \right) ^2 \right \rangle$ for the expression on the left-hand side. Taking the derivative with respect to $s$, we find
\begin{align} \label{stab:gc}
\frac{\partial F}{\partial s}=\sigma^2 \left[
\left\langle \left( \frac{\partial x}{\partial \eta} \right)^2 \right\rangle
+ s \frac{\partial}{\partial s}
\left\langle \left( \frac{\partial x}{\partial \eta} \right)^2 \right\rangle
\right].
\end{align}
The first term in the square bracket captures a direct effect of increasing the number of speculators, assuming the behaviour of all players remains unchanged. It indicates how strongly the activity of any one speculator changes with external perturbations (irrespective of the direction of this change towards higher or lower activity). The term is positive -- speculators are susceptible to perturbations (while producers are not). As $s$ is increased, this has an increasing effect on the left-hand side of the stability condition in (\ref{eq:stab_gc}), and works against stability. 

The second term captures in the bracket in Eq.~(\ref{stab:gc}), $
\frac{\partial}{\partial s}
\left\langle \left( \frac{\partial x}{\partial \eta} \right)^2 \right\rangle$,
captures how the susceptibility of speculators to external perturbations changes when the composition of the player population is changed. This is the result of interactions between different speculators. This term can be negative, and if that is the case, the effects acts to promote stability.

For small $s$, this second term is small compared to the first.  Most speculators are adjusting their strategies to play against the producers, not other speculators, so the first term in Eq.~(\ref{stab:gc}) dominates and the system destabilises with increasing $s$. If $s$ is already large to begin with, the second term can dominate and increasing $s$ further can re-stabilise the system.  
\section{Volatility clustering}
As a final aspect of our analysis we briefly comment on volatility clustering in the SK game. Volatility is known to cluster in time in financial markets (see e.g. \cite{mandelbrot1963variation, mantegna1999introduction}). This means that there are periods of time in which prices move relatively little, and other times where prices move a lot. In the words of Mandelbrot ``large changes tend to be followed by large changes, of either sign, and small changes tend to be followed by small changes" \cite{mandelbrot1963variation}.

\begin{figure}[t!!!]
\centering
     \includegraphics[width=0.8\linewidth]{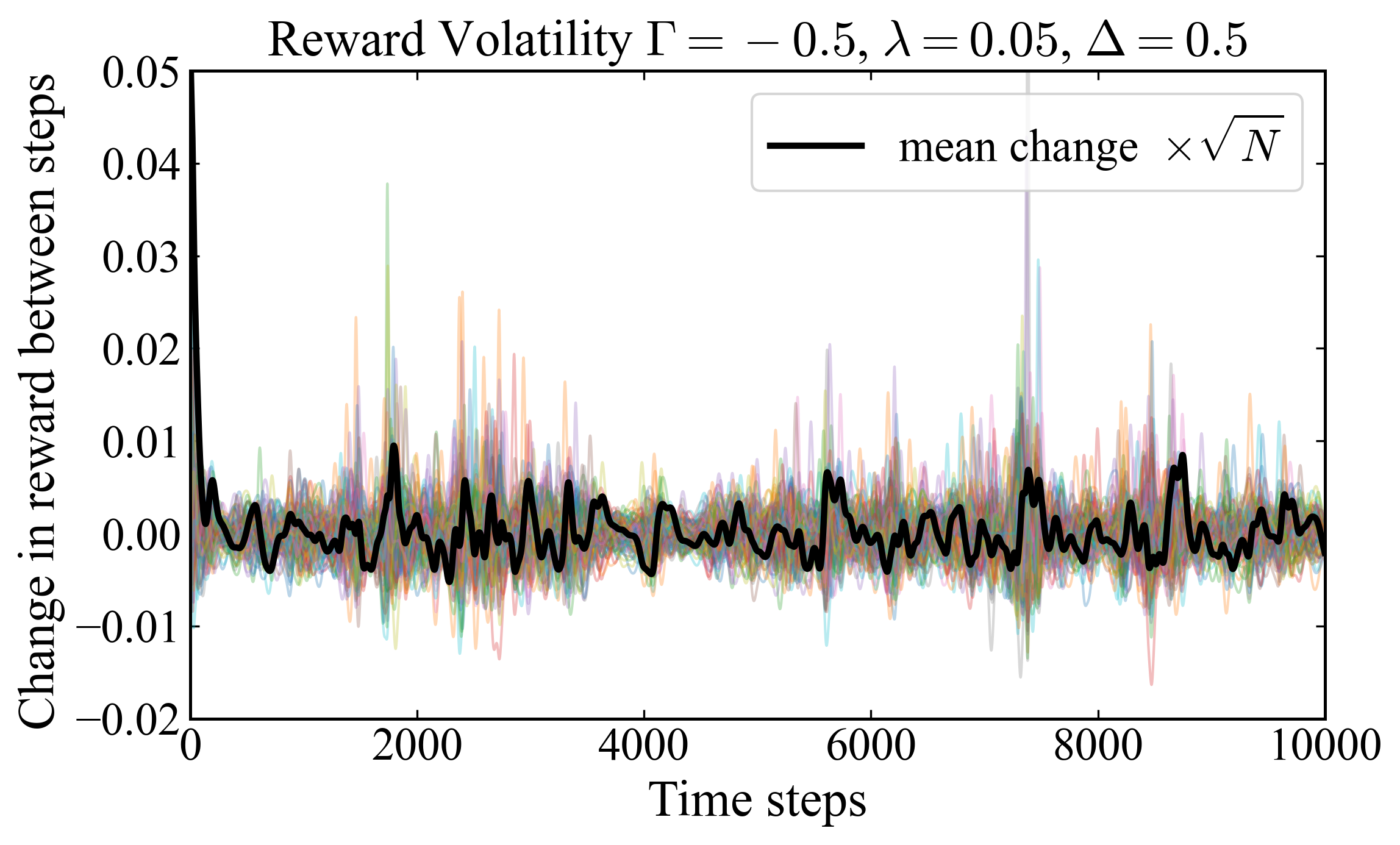}

     \includegraphics[width=0.8\linewidth]{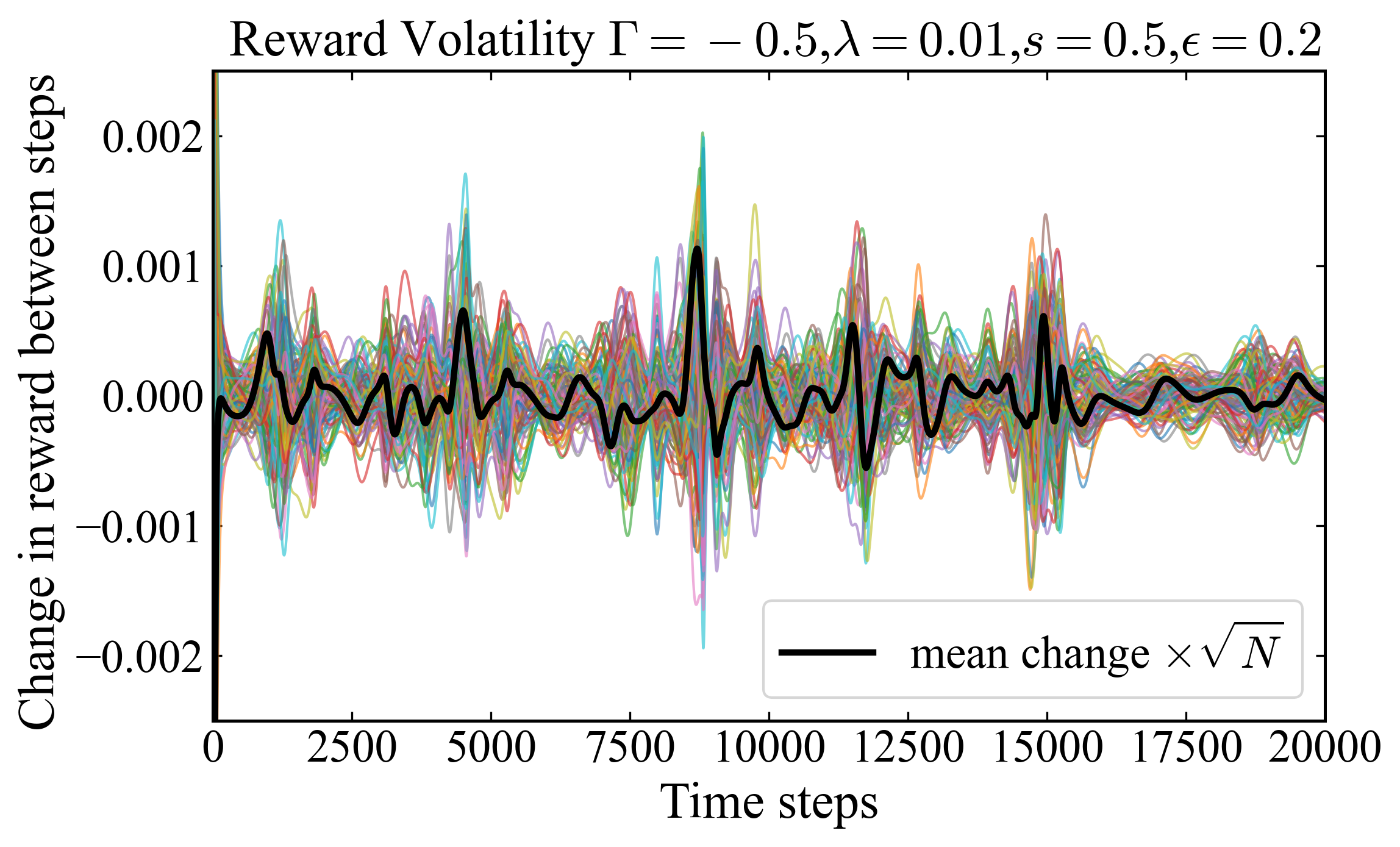}
\caption{Reward volatility, defined as the change in the rewards received by each player between consecutive time steps, is shown for two representative games. The upper panel shows the reward volatility for the SK game with a random field. This corresponds to the game shown in the lower panel of Fig.~\ref{fig:combined2}, simulated over an extended time period. The lower panel shows the reward volatility of the speculator in the grand-canonical SK game. This corresponds to the game shown in the lower panel of Fig.~\ref{fig:combined5}, again simulated over an extended time period.  The black lines shows the mean change in reward, rescaled by $\sqrt{N}$
 so that its fluctuations are comparable in variance to those of a typical agent.}
\label{fig:combined7}
\end{figure}

While the SK game is not a direct model of a market and while payoffs are not an equivalent of prices, we find signs of clustering in simulations of the SK game. To this end, we have measured the change of payoff to individual players $i$ from one round to the next, i.e. $\Pi_i[\bx(t+1)]-\Pi_i[\bx(t)]$. We plot such time series in Fig.~\ref{fig:combined7}. The upper panel is for an instance of the SK game with random external fields in the unstable phase. The lower panel is for a grand-canonical SK game, also in the unstabe phase. In the figure one can clearly see periods of increases payoff movements, interspersed with more quite periods. We also plot a re-scaled version of the mean reward across players, $N^{-1/2}\sum_i \{\Pi_i[\bx(t+1)]-\Pi_i[\bx(t)]\}$. The pre-factor is chosen such that this quantity has fluctuations of the same order as the individidual $\Pi_i[\bx(t+1)]-\Pi_i[\bx(t)]$. Clustered activity can also be seen in this quantity.

\section{Conclusions and discussion}\label{sec:concl}
To summarise, we have studied the outcome of adaptive learning in the Sherrington--Kirkpatrick game. Nash equilibria are the natural outcome for simple games, previous work had shown that adaptive learning does not converge in many complex games with few players but large strategy spaces  \cite{galla2013complex, sanders2018prevalence}. In the Sherrington-Kirkpatrick game, introduced in \cite{garnier2024unlearnable}, a different route to adding complexity is taken. The number of pure strategies available to each player is small (two pure actions), but the number of interacting players is large. In this scenario games also remain `unlearnable', and players have to make do with sub-optimal outcomes.

Our main contributions in this work are to (i) construct the SK game from a concrete payoff structure, (ii) to show that the situation in \cite{garnier2024unlearnable} is a special case in which no player has any intrinsic bias towards any of their two actions, and to implement generalisation to random external fields; (iii) to establish where in parameter space the three different dynamical regimes (unique fixed points, many fixed points, or persistently volatile motion) occur in the SK games with and without field, and (iv) we introduce a `grand-canonical' version of the SK game, in which there are two types of players (speculators and producers), and where speculators have the option to abstain. We establish the stability behaviour for this game, and show that there are interesting re-entrance effects as the fraction of speculator in the population of players is varied.

We think our study is of interest in several different ways. It adds to the body of work on the statistical physics of adaptive agent-learning in complex situations. This programme goes back at least 25 years, when there was significant activity on the Minority Game \cite{challet2006minority, coolen2005mathematical}, and where tools from disordered systems were used to identify a transition between information efficient and inefficient phases. Research, partly pre-dating the MG era addresses similar questions \cite{rieger1989solvable, opper1992phase, berg1, berg2, galla2013complex, sanders2018prevalence} -- What is the typical outcome of complex evolutionary dynamics? -- but focused situations with a large strategy space. The present work, along with that of \cite{garnier2024unlearnable}, instead focuses on scenarios in which many players learn to play Gaussian random games with only two actions per player. It is perhaps surprising that the general phenomenology is rather similar to that of the multi-action two-player games in \cite{galla2013complex}.

This observation makes the work interesting in a more general way. One question that was raised in \cite{galla2013complex, sanders2018prevalence} is if there are general quantifiers which can indicate when a particular learning dynamics would fail to converge for a particular ensemble of games. This question is motivated by observations in fluid dynamics, where the Reynolds number can be used to make approximate predictions as to whether a particular fluid flow can be expected to be turbulent or not. It is now becoming increasingly clear that the degree of competition and the time scale of memory loss in the learning dynamics are key determinants for the stability of learning in a variety of different complex games. There is also increasing evidence that chaos is not an exception, but rather the norm, in different competitive games.
\acknowledgements
	TG thanks Jean-Philippe Bouchaud and J\'er\^ome Garnier-Brun for discussions about their work \cite{garnier2024unlearnable}.
    The work is partially supported by the Agencia Estatal de Investigación and Fondo Europeo de Desarrollo Regional (FEDER, UE) under project COSASTI (PID2024-157493NB-C22), the Mar\'ia de Maeztu programme for Units of Excellence, CEX2021-001164-M, funded by MCIN/AEI/10.13039/501100011033.  
    This work is  also partially supported by a PhD studentship from the Faculty of Natural, Mathematical \& Engineering Sciences, King's College London.

\section*{Data and codes}
Data and codes for the figures available, see \cite{data_and_codes}.

\onecolumngrid
\clearpage

\appendix 

\section{Derivation of Eq.~(\ref{eq:m_learning})}\label{sec:eq_for_m}
For small $\alpha$ and $\beta$, have $
 x^{1-\alpha}e^{\beta \Pi} = x + x (-\alpha \ln\, x +\beta \Pi+\dots)$. We use this expansion in the numerator and in the denominator of the right-hand side of Eq.~(\ref{eq:x_update}), and the fact that $x_i^+(t)+x_i^-(t)=1$. We then arrive at
 \BE
 x_i^\pm(t+1)-x_i^\pm(t) &\approx& x_i^\pm(t) \left\{ -\alpha\ln\,x_i^\pm(t)+\beta\Pi_i^\pm[\bx(t)]-\kappa_i[\bx(t)]\right\},
 \EE
 where $\kappa_i(\bx)=-\alpha (x_i^+\ln\,x_i^+ + x_i^-\ln\,x_i^-)+\beta \overline \Pi_i(\bx)$. Thus, after re-scaling time and writing $\lambda=\alpha/\beta$, we obtain 
  \be\label{eq:ode}
 \frac{d}{dt} x_i^\pm= x_i^\pm \left[ -\lambda\ln\,x_i^\pm(t)+\Pi_i^\pm(\bx)-\nu_i(\bx)\right],
 \ee
 with $ \nu_i(\bx)=-\lambda (x_i^+\ln\,x_i^+ + x_i^-\ln\,x_i^-)+\overline \Pi_i(\bx)$.
\section{Brief comparison with the model of \cite{garnier2024unlearnable}\label{app:compare}}
The learning dynamics in \cite{garnier2024unlearnable} can be summarised as follows [see Eqs. (1), (3) and (5) in \cite{garnier2024unlearnable}],
\be
Q_i(t+1)=(1-\alpha) Q_i(t)+\alpha \sum_j J_{ij} S_j(t),
\ee
where $Q_i$ is the analog of a propensity, and where $S_j$ takes the values $\pm 1$ with probabilities $\frac{1}{2}\left[1\pm \tanh(\beta Q_j(t))\right]$, respectively. The $J_{ij}$ are quenched random couplings.

Taking the batch-limit, we replace $S_j(t)$ with $m_j(t)=\avg{S_j(t)}$. Re-scaling time (to assume that each learning step takes time $\alpha$), and going to a continuous-time limit ($\alpha\ll 1$), we find
\be
\dot Q_i=-Q_i+\sum_j J_{ij} S_j.
\ee
Using Eq.~(2) in \cite{garnier2024unlearnable} and the chain rule, we then find
\be\dot m_i= \beta (1-m_i^2)\left(-Q_i + \sum_j J_{ij} m_j\right).
\ee
We next use $Q_i=\frac{1}{\beta}\mbox{artanh}\,m_i=\frac{1}{2\beta}\ln \frac{1+m_i}{1-m_i}$. We then have
\be\dot m_i= (1-m_i^2)\left(-\frac{1}{2}\ln \frac{1+m_i}{1-m_i} + \beta\sum_j J_{ij} m_j\right).
\ee
\section{Further details of the analysis of the SK game}\label{app:gfsk}
In this appendix we provide details of the generating functional analysis of the learning process of the Sherrington-Kirkpatrick game. This follows the lines of \cite{opper1992phase, galla2013complex}. Further details of this type of calculation can also be found in \cite{galla2024generating}. Alternative approaches based on the cavity method are used (for related Lotka--Volterra systems) in \cite{bunin2017, roy2019numerical}.
\subsection{Generating-functional calculation}
The generating function associated with the SK game with a field from Eq.~(\ref{eq:modified_dyn_m}) is given by
\begin{align} 
    Z(\psi) 
    = 
    \int  \mathcal{D}\mathbf{m}\,\mathcal{D}\hat{\mathbf{m}} \;    
 &\exp\left[i  \sum_i \int dt \hat{m}_i(t)\left( \frac{\dot m_i(t)}{1-m_i(t)^2} + \lambda  \ln\!\left(\frac{1+m_i}{1-m_i}\right) \right)\right]\nonumber \\
\times& \exp\left[i  \sum_i \int dt \hat{m}_i(t)\left( \sum_j A_{ij} m_j(t)+B_{ij}\right) \right] \exp\left(i  \sum_i \int dt m_i(t) \psi_i(t) \right)
\label{eq:gen_fun}
\end{align}
Taking the disorder-average in Eq.~(\ref{eq:gen_fun}), we have:
\begin{align}\label{averaging_step}
    &\overline{ \left[ \exp\left(i  \sum_i \int dt \hat{m}_i(t)\left(\sum_j A_{ij} m_j(t) + B_{ij}\right) \right) \right]} \nonumber\\
    = \; \; \; \;&\exp\left( \frac{\Delta^2}{2}  \sum_i \int dt dt'  \hat{m}_i(t) \hat{m}_i(t') \right  )\times \exp\left( \frac{\gamma\Delta^2}{2}  \sum_{ij} \int dt dt'  \hat{m}_i(t) \hat{m}_j(t') \right  )  \nonumber \\
    \times &\exp\left( \frac{\sigma^2}{2}  \sum_{ij} \int dt dt'  \hat{m}_i(t)  \hat{m}_i(t') \hat{m}_j(t)  \hat{m}_j(t') \right  ) 
    \times \exp\left( \frac{\Gamma \sigma^2}{2}  \sum_{ij} \int dt dt'  \hat{m}_i(t)  \hat{m}_j(t') \hat{m}_i(t')  \hat{m}_j(t) \right  ) \nonumber\\
    = \; \; \; \;& \exp \left( \frac{N}{2} \int dt dt'  \left( \Delta^2  L(t, t')   +  \gamma\Delta^2 P(t) P(t') + \sigma^2 L(t, t') C(t,t')  +  \Gamma \sigma^2 K(t,t') K(t', t)  \right) \right),
\end{align}
where
\begin{align}
P(t) &= \frac{1}{N} \sum_i   \hat{m}_i(t), \nonumber \\
C(t, t') &= \frac{1}{N} \sum_i  m_i(t) m_i(t'),\nonumber \\
K(t, t') &= \frac{1}{N} \sum_i  m_i(t) \hat{m}_i(t'),\nonumber \\ 
L(t, t') &= \frac{1}{N} \sum_i  \hat{m}_i(t) \hat{m}_i(t').
\end{align}
We can now rewrite the disordered averaged generating functional as follows:
\begin{align} \label{eq:saddle}
    \overline{  Z(\Psi) } =  \int \exp(N (\Psi + \Phi + \Omega + O(N^{-1})) \mathcal{D}\left[P, \hat{P},  C , \hat{C}, K, \hat{K} , L , \hat{L} \right],
\end{align}
where
\begin{align}
    \Psi =& \;  i \int dt \hat{P}(t) P(t) + i \int dt dt' \left( \hat{C}(t,t') C(t, t') +\hat{K}(t,t') K(t, t') + \hat{L}(t,t') L(t, t') \right), \nonumber\\
    \Phi =& -\frac{\sigma^2}{2} \int dt dt' \left( L(t,t') C(t,t')  + \Gamma K(t,t') K(t', t)\right) -\frac{\Delta^2}{2}\int dt dt' \left( L(t,t')   + \gamma P(t)P(t')\right), \nonumber\\
    \Omega  = & \; N^{-1} \sum_i 
    \log \int D[m_i , \hat{m}_i]\left[ \exp \left(  i \int dt m_i \psi_i(t)\right)  \right. \nonumber\\
    &\times \exp\left(i  \sum_i \int dt \hat{m}_i(t)\left( \frac{\dot m_i(t)}{1-m_i(t)^2} + \lambda  \ln\!\left(\frac{1+m_i}{1-m_i}\right) \right)\right)\nonumber \\
    & \times \exp\left( -i \int dt dt'  \left[ \hat{C}(t, t') m_i(t)m_i(t') + \hat{L}(t,t') \hat{m}_i(t)\hat{m}_i(t')  +  \hat{K}(t, t') m_i(t)\hat{m}_i(t') \right]\right) \nonumber   \\
    & \left.\times \exp\left( -i \int dt  \hat{P}(t)m(t)\right)  \right]. 
\end{align}
In the limit of $N\to \infty$, we can evaluate the integral in Eq.~(\ref{eq:saddle}) via the saddle-point method. Setting the variation to with respect to the variables $P,  C , K , L $ to zero we obtain:
\begin{align}
i \hat{P}(t) &= \gamma P(t), \nonumber\\
i\hat{C}(t, t') &= \frac{\sigma^2}{2} L(t,t'), \nonumber \\
i\hat{K}(t, t') &= \Gamma \sigma^2 K(t',t), \nonumber \\ 
i\hat{L}(t, t') &=  \frac{1}{2} \left( C(t,t') +  \Delta^2 \right).
\end{align}
Extremising with respect to $\hat{P}$, $\hat C$, $\hat K$, and $\hat L $, we find $P$ and $L = 0$.  This leads to the following effective process
\be
\dot m = -\lambda (1-m^2) \ln \frac{1+m}{1-m} +(1-m^2)\left[\Gamma\sigma^2 \int dt'\, G(t,t') m(t') +\eta(t)\right],
\ee
with the self-consistency relations
\BE
C(t,t')&\equiv& \avg{\eta(t)\eta(t')} = \sigma^2\avg{m(t) m(t')} + \Delta^2 , \nonumber \\
G(t,t')&=& \frac{\delta}{\delta \eta(t')} \avg{m(t)}.
\EE
\subsection{Linear stability analysis for SK game with random field}\label{sec:lin_stab_app}
For clarity we denote quantities evaluated at the fixed point with a superscript $^\star$ in this appendix. We follow the procedure of \cite{opper1992phase}. Linearising about the fixed point involves introducing small perturbations about $m^*$ and $\eta^*$,\begin{align}
    m &\to m^\star + \hat{m}, \nonumber \\
    \eta &\to \eta^\star + \hat{\eta}. 
\end{align}
We obtain the linearised equations about the fixed point:
\begin{align}
    \frac{d}{dt} \hat{m}(t) = -2 \lambda \hat{m} + (1- (m^\star)^2) \left( \Gamma \sigma^2 \int dt'\, G(t - t')\, \hat{m}(t') + \hat{\eta}(t) \right).
\end{align}
We next carrying out a Fourier transform
\begin{align}
    \hat{m}(t) &\to \tilde{m}(\omega), \nonumber\\
    \frac{d}{dt}\hat{m}(t) &\to i\omega \;\tilde{m}(\omega), \nonumber \\
    \hat{\eta}(t) &\to \tilde{\eta}(\omega), \nonumber \\
    \hat{G}(t- t') &\to G(\omega).
\end{align}
We then obtain the following relation
\begin{align}
    \frac{i\omega + 2\lambda - (1 - (m^\star )^2) \Gamma\sigma^2 G(\omega)}{1 - (m^\star )^2 } \; \tilde{m} (\omega)  =  \tilde\eta(\omega)
\end{align}
From this, we find
\begin{align}
    \langle(|\tilde{m} (\omega))|^2 \rangle  &=  \left\langle(|\tilde\eta(\omega))|^2 \left| \frac{(1 - (m^\star )^2) } {i\omega + 2\lambda - (1 - (m^\star )^2) \Gamma\sigma^2 G(\omega) } \right|^2 \right\rangle \\
    &= \frac{1}{\sigma^2} \langle |\tilde\eta(\omega)|^2 \rangle,
\end{align}
where the last identity is obtained from self-consistency relation $\avg{\hat \eta(t)\hat \eta(t')}=\sigma^2\avg{\hat m(t) \hat m(t')}$. %
We note the random field (parametrised by $\Delta$) does not contribute to the linearised dynamics about the fixed point (but the random field does affect the fixed point itself).

As with similar problems \cite{opper1992phase, galla2013complex}, we assume the low frequencies, $\omega \to 0$ are the most unstable.  
In the zero-frequency limit, $\omega \to 0$, perturbing $m^\star$ corresponds to a static displacement of the fixed point.  
Following the lines of \cite{opper1992phase, galla2013complex} we can then conclude that the fixed point is stable when the following relation is fulfilled,
\begin{align}\label{eq: stab_relation}
     \sigma^2 \left\langle \left[ \frac{1 - (m^\star )^2} { 2\lambda - (1 - (m^\star )^2) \Gamma\sigma^2 \chi } \right]^2 \right \rangle = \sigma^2 \left\langle \left( \frac{\partial m^\star } { \partial \eta^\star} \right) ^2 \right \rangle< 1.
\end{align}

\subsection{Case without random field}
When there is no external field ($\Delta=0$), the fixed point distribution becomes concentrated at $m^\star = 0$.  Substituting this into Eq.~\eqref{eq: stab_relation} the stability condition becomes
\begin{align} \label{stab_condition_no_field}
     \sigma  \left[ \frac{1  } { 2\lambda -  \Gamma\sigma^2 \chi } \right] < 1,
\end{align}
which simplifies to
\begin{align} \label{ineq}
     1 < 2\lambda/\sigma -  \Gamma\sigma \chi.
\end{align}
We will now show $\chi = 1/ \sigma$ at the stability boundary.
We note the following relation, valid at the fixed point,
 \be 
-\lambda \ln \frac{1+m^\star}{1-m^\star} + \Gamma\sigma^2 \chi m^\star + \eta^\star =0,
 \ee
Differentiating with respect to $\eta^\star$ we find
 \be  \label{diff_relation}
\left( \frac{ -2 \lambda}{1-(m^\star)^2} + \Gamma\sigma^2 \chi \right) \frac{\partial m^\star}{\partial \eta^\star}+ 1 =0.
 \ee
From this, we obtain
  \be \label{chi_relation}
  \chi= \frac{\partial m^\star}{\partial \eta^\star}= \frac{1 - (m^\star )^2 } {2\lambda - [1 - (m^\star )^2] \Gamma\sigma^2 \chi },
  \ee
  where this is to be evaluated at the fixed point $m^\star=0$. This means that $\chi=(2\lambda-
  \Gamma\sigma^2\chi)^{-1}$. We use this in \eqref{ineq}, and find $\chi = 1/ \sigma$ at the stability boundary.
Substituting this into the relation $\chi=(2\lambda-
  \Gamma\sigma^2\chi)^{-1}$, which we derived earlier,  we find the following condition for the stability boundary,
\be
\lambda_\text{crit} = (1+ \Gamma) \sigma/2.
\ee
Instability occurs for $\lambda < \lambda_\text{crit}$.
\subsection{Approach based on eigenvalues of random matrices}\label{non-dmft}
One can also obtain the stability boundary for the SK-game without random field using an approach based on the spectra of random matrices. This circumvents the DMFT analysis.\\\\
The time evolution of the $\{m_i\}$ is given by
\be\label{eq:dyn_m_app}
 \frac{d}{dt} m_i = (1-m_i^2) \left(\sum_j A_{ij} m_j - \lambda \ln\frac{1+m_i}{1-m_i}\right).
\ee
We now introduce the following variables,
\be \label{change:coor}
y_i = \ln\frac{1+m_i}{1-m_i}
\ee
While the $m_i$ are restricted to the interval $(-1,1)$, the $y_i$ can take values on the entire real axis. This simplifies the analysis. The dynamics for the $y_i$ is given by
\be\label{eq:dyn_c}
 \frac{d}{dt} y_i =2 \left(\sum_j A_{ij} m_j - \lambda y_i\right).
\ee
Expressing the $m_j$ on the right in terms of $y_j$, this can be written in the form
\be\label{eq:dyn_c_2}
 \frac{d}{dt} \mathbf{y} = \mathbf{F}(\mathbf{y}).
\ee
Suppose we are now at a fixed point of the dynamics, written as $m_i= m_i^\star$, i.e., $y_i= y_i^\star$. A perturbation of the vector $\mathbf{y}$ will then evolve as.
\be
 \frac{d}{dt} \delta\mathbf{y} = \underline{\underline{J}} \delta\mathbf{y},
\ee
with a Jacobian matrix, $\underline{\underline{J}}$ given by
\be
J_{ij}
=
\left.\frac{\partial F_i}{\partial y_j}\right|_{\mathbf{y} = \mathbf{y^\star}}
=
\begin{cases}
-2\lambda, & \text{if } i = j, \\
A_{ij} [1- (m^\star_j)^2], & \text{if } i \neq j,
\end{cases}
\ee
In the absence of a random field (i.e., for $\Delta=0$) the system reaches the paramagnetic fixed point, $m_i^\star = 0$ for all $i$. We then have
\be
\underline{\underline{J}} = -2 \lambda \underline{\underline{I}} + \underline{\underline{A}},
\ee
where $\underline{\underline{I}}$ is the identity matrix of size $N\times N$.

We can then determine stability of the paramagnetic fixed point from the eigenvalues of $\underline{\underline{J}}$. The matrix $\underline{\underline{A}}$ is a Gaussian matrix, from Girko's elliptic law \cite{O_Rourke_2015}, we know the eigenvalues (written as $\mu$) of $\underline{\underline{A}}$ (in the $N \to \infty$ limit) are distributed in an ellipse given by:
\be
\frac{(\mbox{Re~} \mu)^2}{ (1+\Gamma)^2}
+
\frac{(\mbox{Im~}\mu)^2}{(1-\Gamma)^2}
\le \sigma^2
\ee
The eigenvalue of $\underline{\underline{J}}$  with the largest real part therefore
\be
\mu_{\mathbf{J},{\rm max}} = \frac{\sigma}{1+ \Gamma} -2 \lambda.
\ee
The fixed point is stable when this quantity is negative, and unstable if it is positive. This leads to the instability condition
\be
\lambda_\text{crit} = (1+ \Gamma) \sigma/2.
\ee
At the onset of instability, the eigenvalue that crosses the imaginary axis has imaginary part equal to zero.    This corresponds to the $\omega \to 0$ mode being the first to destabilise.

\medskip

Introducing a random field $\Delta$ means that the $m_i^\star$ are no longer all zero at the fixed point in the stable phase. In particular the $m_i^\star$ are determined by the matrix $\underline{\underline{A}}$. It is then no longer straightfoward to use existing results on the spectra of random matrices to determine the stability boundary.   However, we note that  $m^\star \neq 0$, weakens the magnitude of the entries in the Jacobian $J_{ij}, i \neq j$, since they are proportional to  $1- (m^\star)^2$.  This means a lower  $\lambda$ would be required to stabilise the system.  This gives an intuition of how introducing $\Delta$ stabilises the system.
\begin{figure}[t]
    \centering
    \includegraphics[width=0.48\linewidth]{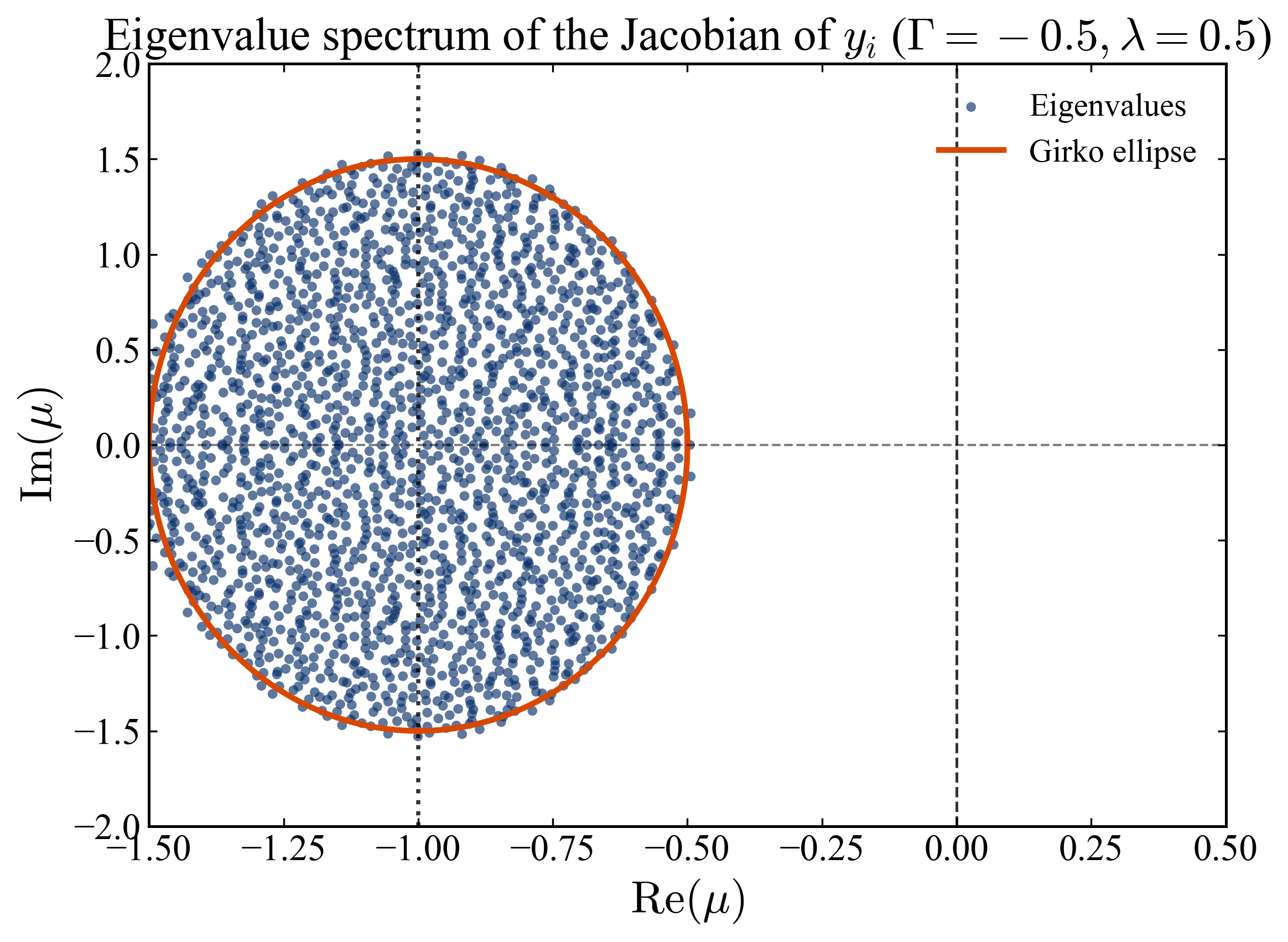}
    \hfill
    \includegraphics[width=0.48\linewidth]{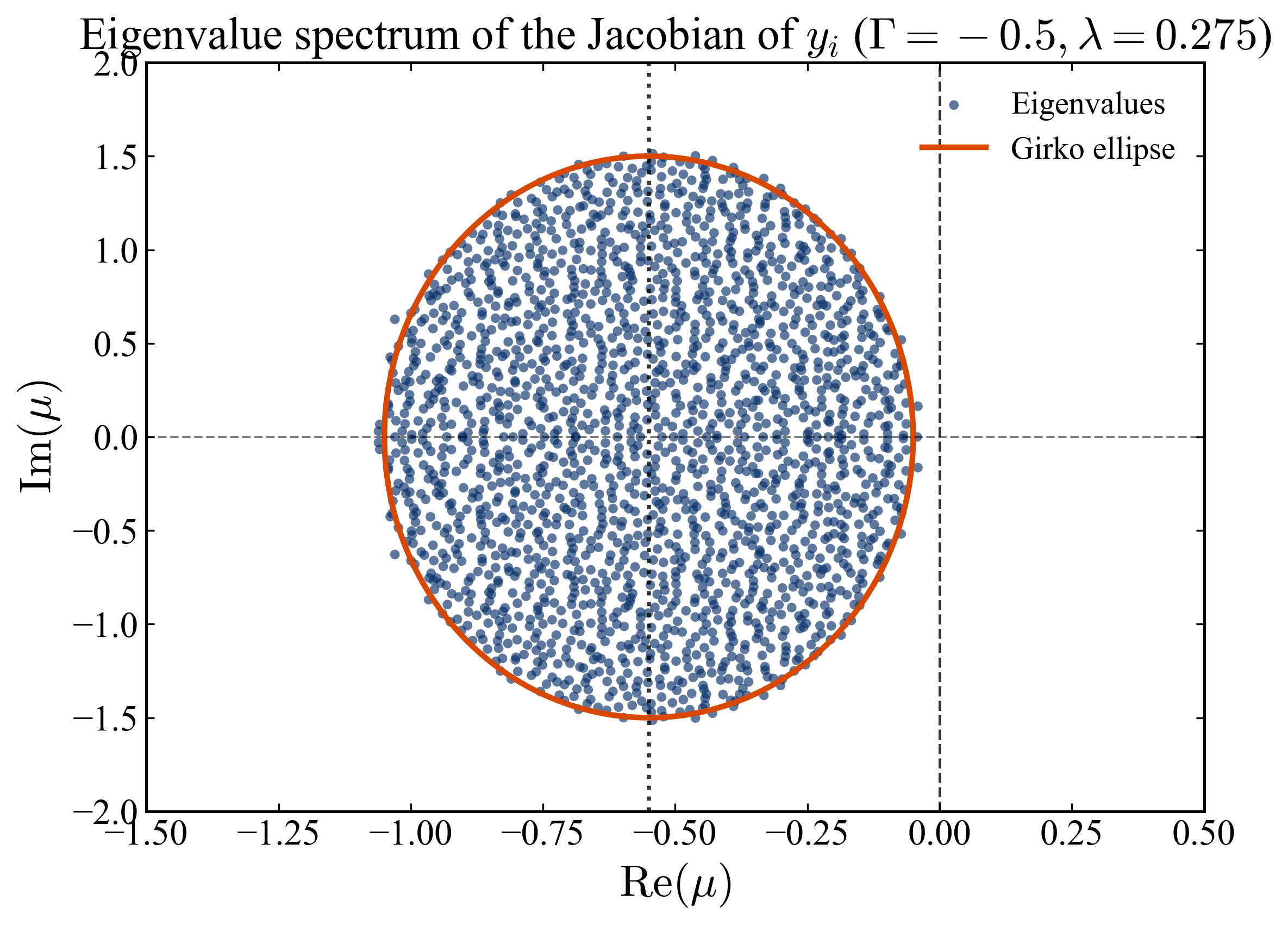}
    \caption{ Plots of the eigenvalues of Jacobian corresponding to $y_i$ at the fixed point for two identical SK games for different values of $\lambda$. The parameters of the game are given by $\Gamma = -0.5, N=2000, \beta= 0.001, \lambda= 0.5, 0.275$, with $\lambda_{crit} = 0.25$.
    Games are run up until convergence to the fixed point.  At the fixed point, we perturb each $y_i$ to compute the corresponding Jacobian.  We plot the eigenvalues of the Jacobian in (blue) against the predicted boundary from random matrix theory (red). 
    These eigenvalues obtained via simulations have been rescaled by a factor of $2$, corresponding to the rescaling of time used in (\ref{eq:dyn_m}). This is necessary because the theoretical calculations are performed in rescaled time.
    We note the real part of the (rescaled) eigenvalues are centered at $-2\lambda$.  }
    \label{eig_[plot]}
\end{figure}
\begin{figure}
    \centering
    \includegraphics[width=0.48\linewidth]{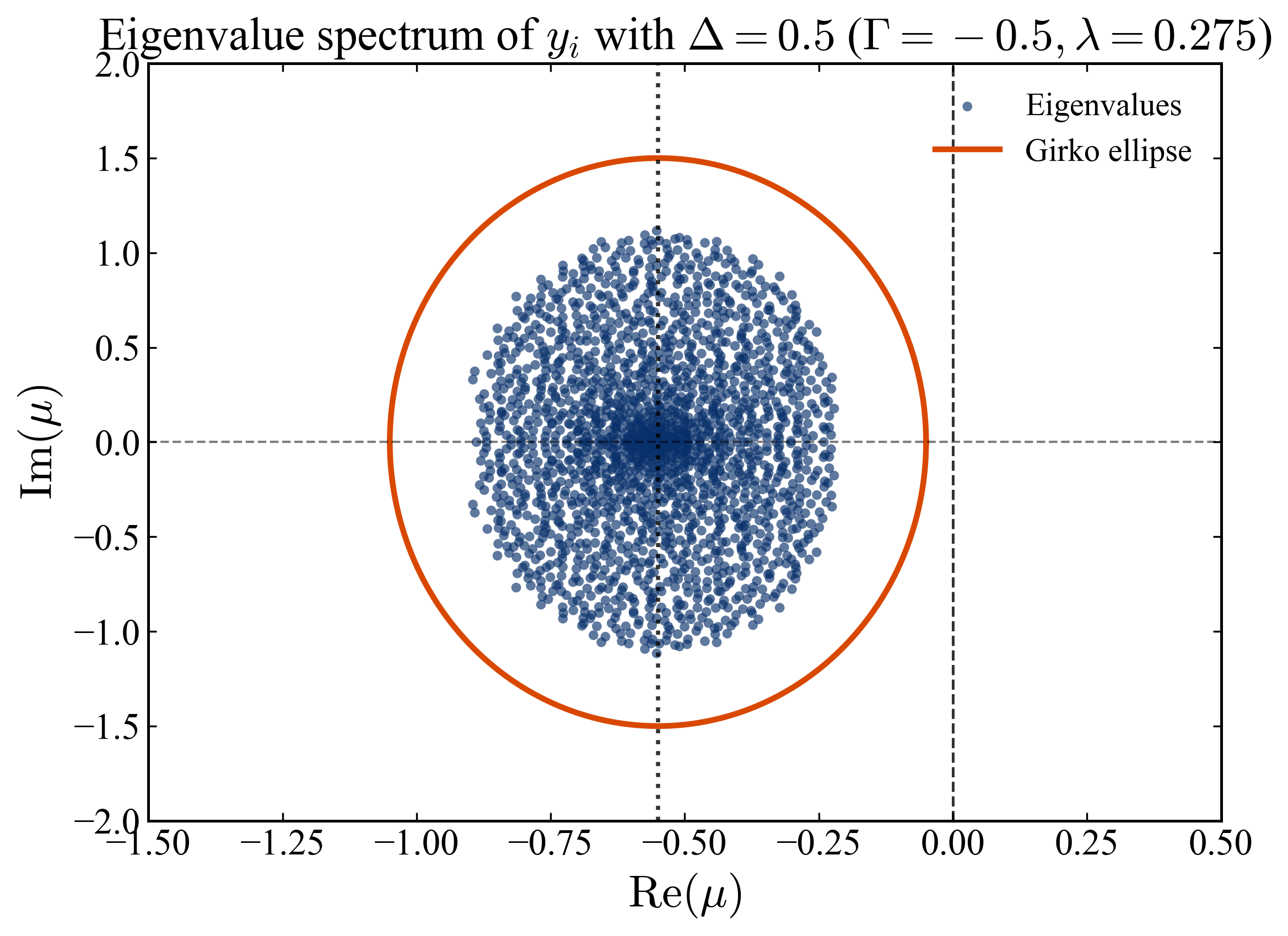}
    \hfill
    \includegraphics[width=0.48\linewidth]{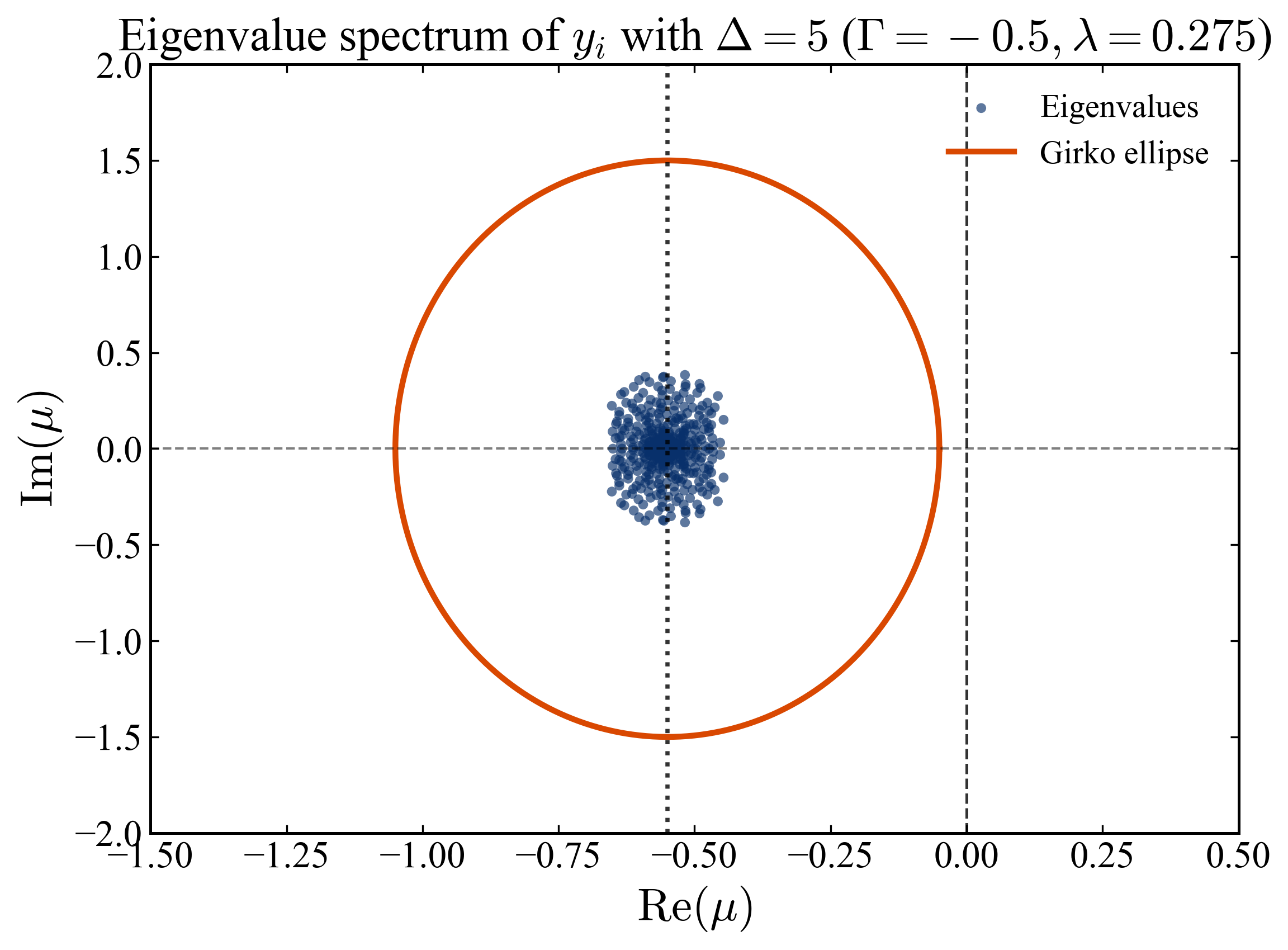}
    \caption{ Plots of the eigenvalues of Jacobian corresponding to $y_i$ for two selected cases with a random field, $\Delta = 0.5, 5$.  The parameters here are similar to the right plot in Figure.~\ref{eig_[plot]}.  We note increasing the strength the random field decreases the how spread of the eigenvalue bulk is.  Since instability occurs when the largest eigenvalue crosses the real axes, increasing the random field requires a lower $\lambda$.}
\end{figure}

\section{Generating-functional calculation for the grand-canonical SK game}\label{app:gcsk_gfa}
\subsection{Dynamic mean-field theory}
The generating functional associated with grand-canonical SK game is given by:
\begin{align} 
    Z(\psi) 
    = 
    \int  \mathcal{D}\mathbf{x}\,\mathcal{D}\hat{\mathbf{x}} \;    
 &\exp\left[i  \sum_i \int dt \hat{x}_i(t)\left( \frac{\dot x_i(t)}{x_i(t)(1-x_i(t))} 
 +
 \lambda  \ln\!\left(\frac{x_i}{1-x_i}\right) \right)\right] \nonumber\\
\times& \exp\left[i  \sum_i \int dt \hat{x}_i(t)\left( \sum_j^{Ns} A_{ij} x_j(t)+h_{i}\right) \right] \times \exp\left(i  \sum_i \int dt x_i(t) \psi_i(t) \right)
\label{eq:gen_fun}
\end{align}
Averaging over the disorder we find
\begin{align}\label{averaging_step}
    &\overline{   \exp\left(i  \sum_i \int dt \hat{x}_i(t)\left( \sum_j^{Ns} A_{ij} x_j(t)+h_{i}\right) \right)  }\nonumber\\
    = \; \; \; \;& \exp \left( \frac{Ns \sigma^2}{2} \int dt dt'  \left(  (1-s) L(t, t')  + s L(t, t') C(t,t')  +  s \Gamma K(t,t') K(t', t)  \right) \right),
\end{align}
where
\begin{align}
C(t, t') &= \frac{1}{Ns} \sum_i  x_i(t) x_i(t'),\nonumber \\
K(t, t') &= \frac{1}{Ns} \sum_i  x_i(t) \hat{x}_i(t'), \nonumber \\ 
L(t, t') &= \frac{1}{Ns} \sum_i  \hat{x}_i(t) \hat{x}_i(t').
\end{align}
At the saddle point we send $Ns \to \infty$. Setting the variation to with respect to  the varibales $C$, $K$, and $L$ to zero we find
\begin{align}
i\hat{C}(t, t') &= \frac{s\sigma^2}{2} L(t,t'), \nonumber\\
i\hat{K}(t, t') &= s \Gamma \sigma^2 K(t',t),\nonumber \\ 
i\hat{L}(t, t') &=  \frac{\sigma^2}{2} \left[s C(t,t') +  (1-s) \right].
\end{align}
This gives the following effective process
\be
\dot x = x (1-x) \left[-\lambda \ln\frac{x}{1-x}+s\Gamma\sigma^2\int dt' G(t,t') x(t')-\varepsilon+\eta(t)\right],
\ee
with
\be
\avg{\eta(t)\eta(t')}= \sigma^2 \left[s \avg{x(t)x(t')}+(1-s)\right].
\ee

\subsection{Linear stability analysis for the grand-canonical game}
We again carry out a linear stability analysis to obtain the stability criterion for the grand-canonical SK game. We find that fixed points are stable when
\begin{align}\label{eq:grand_stab_relation}
     \sigma^2 s\left\langle \left[ \frac{x^\star (1 - x^\star ) } { \lambda - x^\star (1 - x^\star ) s \Gamma\sigma^2 \chi } \right]^2 \right \rangle 
     = 
     \sigma^2 s
     \left\langle  
     \left(\frac{\partial x^\star}{\partial \eta^\star}\right)^2
     \right \rangle < 1.
\end{align}
As shown in Fig.~\ref{fig:gcsk_stab_diagr}, when $\Gamma<0$, increasing the fraction of speculators can destabilise the system when $s$ is small, while for large $s$ increasing $s$ further makes the dynamics more stable. In the following we give some further mathematical reasoning as to why this occurs.

\subsubsection{Increasing the fraction of speculators destabilises learning when $s$ is small}
We introduce the following quantity [see Eq.~(\ref{eq:grand_stab_relation})],
\begin{align}
    F(s)= \sigma^2 s
     \left\langle  
     \left(\frac{\partial x^\star}{\partial \eta^\star}\right)^2
     \right \rangle.
\end{align}
As $F(s)$ increases, the system becomes more unstable, with the phase transition occurring when $F(s) = 1$.  Differentiating with respect to $s$ we find
\begin{align} \label{stab:grandcanonical}
    \frac{\partial F(s)}{\partial s} = \sigma^2 \left[
     \left\langle  
     \left(\frac{\partial x^\star}{\partial \eta^\star}\right)^2
     \right \rangle + s \frac{\partial }{\partial s}\left\langle  
     \left(\frac{\partial x^\star}{\partial \eta^\star}\right)^2
     \right \rangle \right]
\end{align}
The second term is proportional to $s$, and can be neglected for sufficiently small $s$. We then have
\begin{align}
    \left. \frac{\partial F(s)}{\partial s} \right|_{s \to 0} \approx \sigma^2 \left[
     \left\langle  
     \left(\frac{\partial x^\star}{\partial \eta^\star}\right)^2
     \right \rangle \right]>0.
\end{align}
Therefore, for small $s$, increasing $s$ destabilises the system.
\subsubsection{Adding speculators can stabilise when $s$ is already large}
Now we look at the second term in Eq.~(\ref{stab:grandcanonical}). We have
\begin{align} \label{grand_canonical_second_term}
    \frac{\partial }{\partial s}
    \left\langle  
     \left(\frac{\partial x^\star}{\partial \eta^\star}\right)^2 \right \rangle 
     =  
     \frac{\partial }{\partial s}\left\langle \left[ \frac{1 } { \lambda/ (x^\star (1 - x^\star )) -  s \Gamma\sigma^2 \chi } \right]^{2} \right \rangle 
\end{align}
We focus on $\Gamma < 0$. We have: $\lambda/x^\star (1 - x^\star ) >  0$,   and $\Gamma \chi < 0 $ (the susceptibility $\chi$ comes out positive).
We will show that the denominator in \ref{grand_canonical_second_term} increases as $s \to 1$.  Naively, we might assume it is driven by the $-s\Gamma \sigma^2 \chi$ term increasing with $s$, although that might true, in actuality, it is driven by the other term, $\lambda / x^\star (1 - x^\star )$, blowing up.  We will show this in the next segment. \\\\
From Figure \ref{fig:spec_act}, we note $\langle x^\star \rangle$ decreases when $s$ and $\epsilon$ increases. We can get an intuition on how this occurs by looking at the fixed point equation:
\begin{align}
    -\lambda \log\left(\frac{x^\star(z)}{1-x^\star(z)}\right) + s \Gamma \sigma^2  \chi x^\star(z) - \varepsilon +\eta^\star(z) = 0
\end{align}
When $s$ increases, to maintain the equality, $x^\star(z)$ has to decrease.  Likewise, when $\varepsilon$ increases. Thus, for some sufficiently large epsilon $\epsilon > \epsilon_0$, as we increase $s \to 1$, we effectively have $\langle x^\star \rangle \to 0$, which causes the denominator in \ref{grand_canonical_second_term} to blow up.
Therefore, $ \frac{\partial }{\partial s}\left\langle  
     \left(\frac{\partial x^\star}{\partial \eta^\star}\right)^2 \right \rangle$ is  negative for large $s$, and as  $s$ increases, we are placing more `weight' on the strictly negative term in \ref{stab:grandcanonical}:
\begin{align} \label{stab:grandcanonical2}
\frac{\partial F(s)}{\partial s}
&=
\sigma^2 \left[
\underbrace{
\left\langle  
\left(\frac{\partial x^\star}{\partial \eta^\star}\right)^2
\right\rangle
}_{>0 (\text{dominant at small s})}
+
\underbrace{
s \frac{\partial }{\partial s}\left\langle  
\left(\frac{\partial x^\star}{\partial \eta^\star}\right)^2
\right\rangle
}_{<0 (\text{can become dominant at larger s})}
\right].
\end{align}
Thus, increasing $s$ can decrease $F(s)$ and therefore stabilise the system.

\bibliography{references} 

@misc{data_and_codes,
  title = {Data and codes: },
  howpublished = {\url{https://github.com/Desmondccw/complex_dynamics_in_sk_game}},
}

@book{mantegna1999introduction,
  title={Introduction to econophysics: correlations and complexity in finance},
  author={Mantegna, Rosario N and Stanley, H Eugene},
  year={1999},
  publisher={Cambridge University Press}
}

@article{traulsen2009stochastic,
  title={Stochastic evolutionary game dynamics},
  author={Traulsen, Arne and Hauert, Christoph},
  journal={Reviews of Nonlinear Dynamics and Complexity},
  volume={2},
  pages={25--61},
  year={2009},
  publisher={Wiley Online Library}
}

@article{hauert2005game,
  title={Game theory and physics},
  author={Hauert, Christoph and Szab{\'o}, Gy{\"o}rgy},
  journal={American Journal of Physics},
  volume={73},
  number={5},
  pages={405--414},
  year={2005},
  publisher={American Association of Physics Teachers}
}

@article{pangallo2019best,
  title={Best reply structure and equilibrium convergence in generic games},
  author={Pangallo, Marco and Heinrich, Torsten and Doyne Farmer, J},
  journal={Science Advances},
  volume={5},
  number={2},
  pages={eaat1328},
  year={2019},
  publisher={American Association for the Advancement of Science}
}

@article{
Nash,
author = {John F. Nash },
title = {Equilibrium points in <i>n</i>-person games},
journal = {Proceedings of the National Academy of Sciences},
volume = {36},
number = {1},
pages = {48-49},
year = {1950}}

@book{vonNeumann1944,
  title     = {Theory of Games and Economic Behavior},
  author    = {von Neumann, John and Morgenstern, Oskar},
  year      = {1944},
  publisher = {Princeton University Press},
  address   = {Princeton, NJ}
}

@article{rieger1989solvable,
  title={Solvable model of a complex ecosystem with randomly interacting species},
  author={Rieger, H},
  journal={Journal of Physics A: Mathematical and General},
  volume={22},
  number={17},
  pages={3447--3460},
  year={1989}
}

@article{mandelbrot1963variation,
  title={The variation of certain speculative prices},
  author={Mandelbrot, Benoit and others},
  journal={Journal of Business},
  volume={36},
  number={4},
  pages={394},
  year={1963},
  publisher={Springer}
}

@article{sidhom,
  title = {Ecological communities from random generalized Lotka-Volterra dynamics with nonlinear feedback},
  author = {Sidhom, Laura and Galla, Tobias},
  journal = {Phys. Rev. E},
  volume = {101},
  issue = {3},
  pages = {032101},
  numpages = {15},
  year = {2020},
  month = {Mar},
  publisher = {American Physical Society},
}

@article{bunin2017,
  title = {Ecological communities with Lotka-Volterra dynamics},
  author = {Bunin, Guy},
  journal = {Phys. Rev. E},
  volume = {95},
  issue = {4},
  pages = {042414},
  numpages = {8},
  year = {2017},
  month = {Apr},
  publisher = {American Physical Society},
}

@article{de_dominicis,
  title = {Dynamics as a substitute for replicas in systems with quenched random impurities},
  author = {De Dominicis, C.},
  journal = {Phys. Rev. B},
  volume = {18},
  issue = {9},
  pages = {4913--4919},
  numpages = {0},
  year = {1978},
  month = {Nov},
  publisher = {American Physical Society},
}

@book{coolen2005mathematical,
  title={The Mathematical Theory of Minority Games: Statistical Mechanics of Interacting Agents},
  author={{A.~C.~C. Coolen}},
  year={2005},
  publisher={Oxford University Press, Oxford UK}
}

@article{galla2024generating,
  title={Generating-functional analysis of random Lotka-Volterra systems: A step-by-step guide},
  author={Galla, Tobias},
  journal={arXiv preprint arXiv:2405.14289},
  year={2024}
}

@article{Traulsen2004PRL,
  title = {Stochastic Gain in Population Dynamics},
  author = {Traulsen, Arne and R\"ohl, Torsten and Schuster, Heinz Georg},
  journal = {Phys. Rev. Lett.},
  volume = {93},
  issue = {2},
  pages = {028701},
  numpages = {4},
  year = {2004},
  month = {Jul},
  publisher = {American Physical Society},
}

@article{may1972will,
  title={Will a large complex system be stable?},
  author={May, Robert M},
  journal={Nature},
  volume={238},
  number={5364},
  pages={413--414},
  year={1972},
  publisher={Nature Publishing Group UK London}
}

@article{berg1,
  title = {Matrix Games, Mixed Strategies, and Statistical Mechanics},
  author = {Berg, J. and Engel, A.},
  journal = {Phys. Rev. Lett.},
  volume = {81},
  issue = {22},
  pages = {4999--5002},
  numpages = {0},
  year = {1998},
  month = {Nov},
  publisher = {American Physical Society},
}

@article{berg2,
year = {1999},
month = {oct},
publisher = {},
volume = {48},
number = {2},
pages = {129},
author = {J. Berg and M. Weigt},
title = {Entropy and typical properties of Nash equilibria in 
two-player games},
journal = {Europhysics Letters},
}

@article{mclennan,
title = {Asymptotic expected number of Nash equilibria of two-player normal form games},
journal = {Games and Economic Behavior},
volume = {51},
number = {2},
pages = {264-295},
year = {2005},
note = {Special Issue in Honor of Richard D. McKelvey},
author = {Andrew McLennan and Johannes Berg},
}

@article{daskalakis, author = {Daskalakis, Constantinos and Goldberg, Paul W. and Papadimitriou, Christos H.}, title = {The complexity of computing a Nash equilibrium}, year = {2009}, issue_date = {February 2009}, publisher = {Association for Computing Machinery}, address = {New York, NY, USA}, volume = {52}, number = {2}, journal = {Commun. ACM}, month = feb, pages = {89–97}, numpages = {9} }

@article{sato2,
  title = {Coupled replicator equations for the dynamics of learning in multiagent systems},
  author = {Sato, Yuzuru and Crutchfield, James P.},
  journal = {Phys. Rev. E},
  volume = {67},
  issue = {1},
  pages = {015206(R)},
  numpages = {4},
  year = {2003},
  month = {Jan},
  publisher = {American Physical Society},
}

@article{sato1,
author = {Yuzuru Sato  and Eizo Akiyama  and J. Doyne Farmer },
title = {Chaos in learning a simple two-person game},
journal = {Proceedings of the National Academy of Sciences},
volume = {99},
number = {7},
pages = {4748-4751},
year = {2002}}

@article{sato3,
title = {Stability and diversity in collective adaptation},
journal = {Physica D: Nonlinear Phenomena},
volume = {210},
number = {1},
pages = {21-57},
year = {2005},
author = {Yuzuru Sato and Eizo Akiyama and James P. Crutchfield}
}

@article{galla2013complex,
  title={Complex dynamics in learning complicated games},
  author={Galla, Tobias and Farmer, J Doyne},
  journal={Proceedings of the National Academy of Sciences},
  volume={110},
  number={4},
  pages={1232--1236},
  year={2013},
  publisher={National Academy of Sciences}
}

@article{garnier2024unlearnable,
  title={Unlearnable games and “satisficing” decisions: a simple model for a complex world},
  author={Garnier-Brun, Jerome and Benzaquen, Michael and Bouchaud, Jean-Philippe},
  journal={Physical Review X},
  volume={14},
  number={2},
  pages={021039},
  year={2024},
  publisher={APS}
}

@article{sanders2018prevalence,
  title={The prevalence of chaotic dynamics in games with many players},
  author={Sanders, James BT and Farmer, J Doyne and Galla, Tobias},
  journal={Scientific Reports},
  volume={8},
  number={1},
  pages={4902},
  year={2018},
  publisher={Nature Publishing Group UK London}
}

@article{challet2006minority,
  title={Minority games with finite score memory},
  author={Challet, Damin and De Martino, Andrea and Marsili, Matteo and Perez Castillo, Isaac},
  journal={Journal of Statistical Mechanics: Theory and Experiment},
  volume={2006},
  number={03},
  pages={P03004--P03004},
  year={2006}
}

@article{menezes2013minority,
  title={Minority game with SK interactions},
  author={Menezes, Pedro Castro and Sherrington, David},
  journal={Journal of Physics A: Mathematical and Theoretical},
  volume={46},
  number={50},
  pages={505004},
  year={2013},
  publisher={IOP Publishing}
}

@article{camerer1999experience,
  title={Experience-weighted attraction learning in normal form games},
  author={Camerer, Colin and Hua, Teck-Ho},
  journal={Econometrica},
  volume={67},
  number={4},
  pages={827--874},
  year={1999},
  publisher={Wiley Online Library}
}

@article{opper1992phase,
  title={Phase transition and 1/f noise in a game dynamical model},
  author={Opper, Manfred and Diederich, Sigurd},
  journal={Physical Review Letters},
  volume={69},
  number={10},
  pages={1616},
  year={1992},
  publisher={APS}
}

@article{roy2019numerical,
  title={Numerical implementation of dynamical mean field theory for disordered systems: Application to the Lotka--Volterra model of ecosystems},
  author={Roy, Felix and Biroli, Giulio and Bunin, Guy and Cammarota, Chiara},
  journal={Journal of Physics A: Mathematical and Theoretical},
  volume={52},
  number={48},
  pages={484001},
  year={2019},
  publisher={IOP Publishing}
}

@article{sherrington1975solvable,
  title={Solvable model of a spin-glass},
  author={Sherrington, David and Kirkpatrick, Scott},
  journal={Physical Review Letters},
  volume={35},
  number={26},
  pages={1792},
  year={1975},
  publisher={APS}
}

@article{challet2000modeling,
  title={Modeling market mechanism with minority game},
  author={Challet, Damien and Marsili, Matteo and Zhang, Yi-Cheng},
  journal={Physica A: Statistical Mechanics and its Applications},
  volume={276},
  number={1-2},
  pages={284--315},
  year={2000},
  publisher={Elsevier}
}

@article{O_Rourke_2015,
   title={Products of Independent Elliptic Random Matrices},
   volume={160},
   number={1},
   journal={Journal of Statistical Physics},
   publisher={Springer Science and Business Media LLC},
   author={O’Rourke, Sean and Renfrew, David and Soshnikov, Alexander and Vu, Van},
   year={2015},
   month=Apr, pages={89–119} }
 
\end{document}